%% file: Paper-DeltaB.tex
\documentclass[a4paper, 10pt]{article}

\usepackage{jheppub}

\usepackage{amsthm}

\usepackage[numbers,sort&compress]{natbib}
\usepackage{dsfont}
\usepackage{slashed}
\usepackage{color}
\usepackage{empheq}

\usepackage{adjustbox}

\usepackage{bibgerm}

\usepackage{soul}

\usepackage[english]{babel}

\usepackage[utf8x]{inputenc}

\usepackage{bashful}

\usepackage{subfigure}

\usepackage{setspace}

\usepackage{booktabs}

\usepackage{float}

\usepackage{nextpage}

\usepackage{pdfpages}

\usepackage{hypcap}

\usepackage[small,bf]{caption}

\usepackage{longtable}

\usepackage{fancyhdr}

\usepackage[makeroom]{cancel}

\usepackage{microtype}

\usepackage{xparse}
\usepackage{etoolbox}

\usepackage[subfigure]{tocloft}

\newcommand{\p}{\partial}

\renewcommand{\O}{\mathcal{O}}

\renewcommand{\L}{\mathcal{L}}

\newcommand{\E}{\mathcal{E}}
\newcommand{\nn}{\nonumber\\}

\newcommand{\q}{\mathsf{q}}

\newcommand{\hc}{\mathrm{h.c.}}

\newcommand{\op}[3]{\O^{#2,#3}_{#1}}

\NewDocumentCommand{\Op}{ m m O{} o }{
	\O^{\ifblank{#3}{}{#3,}#2 }_{\IfNoValueTF{#4}{#1}{\substack{#1\\#4}}}
}
\NewDocumentCommand{\EOp}{ m m O{} o }{
	\E^{\ifblank{#3}{}{#3,}#2 }_{\IfNoValueTF{#4}{#1}{\substack{#1\\#4}}}
}
\NewDocumentCommand{\lwc}{ m m O{} o }{
	L^{\ifblank{#3}{}{#3,}#2 }_{\IfNoValueTF{#4}{#1}{\substack{#1\\#4}}}
}
\NewDocumentCommand{\kwc}{ m m O{} o }{
	K^{\ifblank{#3}{}{#3,}#2 }_{\IfNoValueTF{#4}{#1}{\substack{#1\\#4}}}
}
\NewDocumentCommand{\dlwc}{ m m O{} o }{
	{\dot L}^{\ifblank{#3}{}{#3,}#2 }_{\IfNoValueTF{#4}{#1}{\substack{#1\\#4}}}
}

\makeatletter
  \def\my@tag@font{\normalsize}
  \def\maketag@@@#1{\hbox{\m@th\normalfont\my@tag@font#1}}
  \let\amsmath@eqref\eqref
  \renewcommand\eqref[1]{{\let\my@tag@font\relax\amsmath@eqref{#1}}}
\makeatother

\makeatletter
\renewcommand\paragraph{\@startsection{paragraph}{4}{\z@}%
  {-3.25ex\@plus -1ex \@minus -.2ex}%
  {1.5ex \@plus .2ex}%
  {\normalfont\normalsize\bfseries}}
\makeatother

\cftsetindents{section}{0em}{2em}
\cftsetindents{subsection}{2em}{3em}
\cftsetindents{subsubsection}{5em}{4em}

\allowdisplaybreaks[1]

\preprint{
\mbox{}\hfill{} PSI-PR-25-09 \\
\mbox{}\hfill{} ZU-TH 31/25 \\
\mbox{}\hfill{} INT-PUB-25-013
}

\title{\boldmath Renormalization-group equations of the LEFT at two loops: dimension-six baryon-number-violating operators}

\author{Luca Naterop,}

\author{Peter Stoffer}

\emailAdd{luca.naterop@physik.uzh.ch}
\emailAdd{stoffer@physik.uzh.ch}

\affiliation{Physik-Institut, Universit\"at Z\"urich, Winterthurerstrasse 190, 8057 Z\"urich, Switzerland}
\affiliation{PSI Center for Neutron and Muon Sciences, 5232 Villigen PSI, Switzerland}

\abstract{
	We present the second part of a systematic calculation of the two-loop anomalous dimensions for the low-energy effective field theory below the electroweak scale (LEFT): the baryon-number-violating sector at dimension six in the power counting. We obtain the results in two different schemes: in the algebraically consistent 't Hooft--Veltman scheme for $\gamma_5$, corrected for evanescent as well as chiral-symmetry-breaking effects through finite renormalizations; and in naive dimensional regularization, which in the considered sector of the theory does not lead to any ill-defined $\gamma_5$-odd traces. Our results are of interest for a reanalysis of the constraints on physics beyond the Standard Model from proton-decay searches within an EFT framework at next-to-leading-logarithmic accuracy.
}

\numberwithin{equation}{section}

\begin{document}

	\maketitle

		
	\input{sections/Introduction}
	\input{sections/LEFT}

	\input{sections/Flavor}
	\input{sections/NDR1Loop}
	\input{sections/Computation}
	\input{sections/Results}
	\input{sections/Conclusions}
	
	\section*{Acknowledgements}
	\addcontentsline{toc}{section}{\numberline{}Acknowledgements}

	We thank S.~Banik, A.~Crivellin, B.~Grinstein, A.~V.~Manohar, B. Ruijl, C.-H.~Shen, A.~Signer, D.~St\"ockinger, A.~E.~Thomsen, and M.~Zoller for useful discussions
	and P.~Morell and M.~Pesut for helpful exchange about Ref.~\cite{Aebischer:2025hsx}.
	LN thanks the Institute for Nuclear Theory at the University of Washington for its kind hospitality and stimulating research environment. This research was supported in part by the INT's U.S.\ Department of Energy grant No.~DE-FG02-00ER41132.
	Financial support by the Swiss National Science Foundation (Project No.~PCEFP2\_194272) is gratefully acknowledged.

	
	\appendix
	
	\input{sections/RGE}

	\addcontentsline{toc}{section}{\numberline{}References}
	\bibliographystyle{utphysmod}
	\bibliography{Literature}
	
\end{document}

%% file: sections/Introduction.tex

\section{Introduction}

Our universe contains far more baryons than antibaryons. The present-time baryon asymmetry, as obtained from measurements of the cosmic microwave background and big-bang nucleosynthesis theory, is $\eta = (n_B - n_{\bar{B}})/n_\gamma  = (6.12 \pm 0.04) \times 10^{-10}$~\cite{ParticleDataGroup:2024cfk,Planck:2018vyg,Yeh:2022heq}. Even though the Standard Model (SM) violates baryon number $B$ non-perturbatively \cite{tHooft:1976rip,tHooft:1976snw} (while conserving $B-L$), SM baryogenesis faces the problem that the electroweak phase transition is not of first order~\cite{Kajantie:1996mn,Gurtler:1997hr,Laine:1998jb,Csikor:1998eu,Aoki:1999fi}. The baryon asymmetry is thus thought to have arisen dynamically from $B$-violating effects beyond the SM, as codified in one of Sakharov's conditions \cite{Sakharov:1967dj}. Many experimental and theoretical efforts to search for baryon-number violation are under way~\cite{Babu:2013jba,Broussard:2025opd}. Prominent examples of scenarios beyond the SM that violate $B$ are Grand Unified Theories (GUTs)~\cite{Pati:1973uk,Georgi:1974sy,Fritzsch:1974nn}, which unify all elementary particle forces, or theories of leptogenesis. It is also expected that quantum gravity violates $B$ \cite{ZELDOVICH1976254,HAWKING1979175,Kallosh:1995hi}. 

The proton is known to have a lifetime of at least $10^{34}$ years~\cite{Super-Kamiokande:2014otb,Super-Kamiokande:2020wjk}, and $B$-violating scenarios are tightly constrained by bounds on a variety of decay channels~\cite{ParticleDataGroup:2024cfk}. Within an effective field theory (EFT) framework~\cite{Wilczek:1979hc,Weinberg:1979sa,Abbott:1980zj,Buchmuller:1985jz,Grzadkowski:2010es,Alonso:2014zka}, proton-decay searches yield bounds on $\Delta B \neq 0$ interactions starting at dimension six~\cite{Hou:2005iu,deGouvea:2014lva,Heeck:2019kgr,Beneito:2023xbk,Crivellin:2023ter,Beneke:2024hox,Heeck:2024jei,Gisbert:2024sjw}. Other possible low-energy probes are baryon--antibaryon oscillations or two-nucleon decays, which are described by operators of mass dimension 9 and 12, respectively~\cite{Buchoff:2015qwa,He:2021sbl,He:2021mrt}. Given the large separation of scales involved in $B$-violation searches, EFT methods are crucial to improve perturbation theory through resummation of large logarithms. 

A full analysis makes use of a tower of EFTs, in particular the SMEFT above the electroweak scale~\cite{Buchmuller:1985jz,Grzadkowski:2010es} and the low-energy effective field theory (LEFT) below the electroweak scale~\cite{Jenkins:2017jig}. The dimension-six one-loop anomalous dimensions~\cite{Abbott:1980zj,Jenkins:2013zja,Jenkins:2013wua,Alonso:2013hga,Alonso:2014zka,Jenkins:2017dyc} in these theories and the one-loop matching between SMEFT and LEFT~\cite{Dekens:2019ept} are known. An EFT framework at next-to-leading-logarithmic accuracy requires knowledge of the two-loop anomalous dimensions. In particular, scheme dependences in the one-loop matching have to cancel against similar ones in the two-loop anomalous dimensions. In this paper, we calculate the $\Delta B \neq 0$ two-loop anomalous dimensions in the LEFT at dimension six in the power counting. This represents a further step in a broader program towards the complete two-loop anomalous dimensions in the LEFT and SMEFT~\cite{Gorbahn:2016uoy,deVries:2019nsu,Bern:2020ikv,Aebischer:2022anv,Fuentes-Martin:2022vvu,Aebischer:2023djt,Jenkins:2023rtg,Jenkins:2023bls,Naterop:2023dek,DiNoi:2023ygk,Fuentes-Martin:2023ljp,Aebischer:2024xnf,Manohar:2024xbh,DiNoi:2024ajj,Born:2024mgz,Naterop:2024ydo,Fuentes-Martin:2024agf,Aebischer:2025hsx,Duhr:2025zqw,Haisch:2025lvd}. We perform the calculation in two different schemes: the first one is the scheme that we established for the LEFT in Ref.~\cite{Naterop:2023dek}, the 't~Hooft--Veltman (HV) scheme for $\gamma_5$~\cite{tHooft:1972tcz,Breitenlohner:1977hr} (sometimes called BMHV scheme), supplemented by finite renormalizations that restore chiral symmetry (in the spurion sense) and compensate evanescent-operator insertions. The HV scheme is the only scheme known to be algebraically consistent to all loop orders~\cite{Breitenlohner:1975hg,Breitenlohner:1976te,Breitenlohner:1977hr,Schubert:1988ke}. These results are part of a complete calculation of the two-loop renormalization-group equations (RGEs) of the LEFT in the rigorous HV scheme, which we started in Ref.~\cite{Naterop:2024ydo}. The second scheme that we use is naive dimensional regularization (NDR), i.e., a fully anticommuting $\gamma_5$: in the two-loop renormalization of the $\Delta B \neq 0$ sector of the LEFT, no problematic $\gamma_5$-odd traces appear and hence NDR can be consistently used.

In Sect.~\ref{sec:LEFT}, we define our conventions for the LEFT in the two schemes. In particular, we list the operators including evanescent ones, which are needed in dimensional regularization for a proper scheme definition. In this context, we explain some subtleties that are specific to the $\Delta B \neq 0$ sector. In Sect.~\ref{sec:Flavor}, we discuss the flavor structure and symmetries of the involved LEFT operators. We present the one-loop renormalization in NDR in Sect.~\ref{sec:OneLoopNDR} and explain our methods for the two-loop calculation in Sect.~\ref{sec:Computation}. In Sect.~\ref{sec:Results}, we discuss the resulting RGEs both in the HV and the NDR schemes. In particular, we compare the NDR results to existing literature~\cite{Nihei:1994tx,Gracey:2012gx,Aebischer:2025hsx}, finding partial agreement. We conclude in Sect.~\ref{sec:Conclusions}. The explicit results for the RGEs are provided in App.~\ref{sec:RGEs} for an arbitrary number of flavors.

%% file: sections/LEFT.tex

\section{Baryon-number-violating operators in the LEFT}
\label{sec:LEFT}

\subsection{Physical sector}

The dimension-six basis of the LEFT~\cite{Jenkins:2017jig} contains 16 operator types that violate baryon number $B$, reproduced in Tab.~\ref{tab:PhysicalOperators}. At this dimension, the $\Delta B$ sector does not mix with any other sector of the LEFT, and thus the RGEs can be derived without consideration of the other sectors. As we are only interested in the mixing of dimension-six operators without derivatives, the calculation is not sensitive to mass terms of the theory. The Lagrangian is then\footnote{For the calculations in this sector, at dimension six it suffices to include neutrinos only as external fields in the operators.}
\begin{equation}
    \label{eq:LEFTLagrangian}
    \L_\mathrm{LEFT} = \L_\mathrm{QCD+QED} + \sum_i L_i \O_i + \sum_i K_i \E_i \, ,
\end{equation}
with the dimension-four terms  
\begin{align}
	\label{eq:qcdqed}
	\L_{\rm QCD + QED} \; = \; - \frac14 G_{\mu \nu}^A G^{A \mu \nu} -\frac14 F_{\mu \nu} F^{\mu\nu}  + \sum_{\psi=u,d,e}\overline \psi i \slashed D \psi \;+\; \L_\text{gauge-fixing} \;+\; \L_\mathrm{ghosts}
\end{align}
and covariant derivative $D_\mu = \p_\mu + i g T^A G_\mu^A + i e Q A_\mu$. In addition to the physical operators $\O_i$ from Tab.~\ref{tab:PhysicalOperators}, in dimensional regularization we also encounter evanescent operators $\E_i$ with coefficients $K_i$. Even though they vanish in four dimensions, the $\E_i$ affect the renormalization group, starting at the two-loop order. The precise definition of the $\mathcal{E}_i$ depends on the choice of scheme. In particular, we require a different set of evanescent operators in the HV and NDR schemes for $\gamma_5$.

\input{include/operatorsphysical.tex}

\subsection{Evanescent operators in the HV scheme}

In the HV scheme, we split space-time into a four-dimensional part, with Lorentz indices \mbox{$\bar{\mu} \in \{ 0,1,2,3 \}$}, and an extra-dimensional part, with Lorentz indices $\hat{\mu}$ covering the fractional $D-4 = -2\varepsilon$ dimensions. The prescription for $\gamma_5$ is
\begin{equation}
	\{\gamma_5,\bar\gamma_\mu\} = 0 \quad \text{and} \quad [\gamma_5,\hat\gamma_\mu] = 0 \, ,
\end{equation}
which induces a split between the four-dimensional and extra-dimensional part of space-time. As a result, up to dimension six we encounter two types of evanescent operators: the first operator type explicitly involves contractions restricted to evanescent indices. The second type are Fierz-evanescent four-fermion operators, since away from four-dimensional spacetime, the Fierz relations do not even hold for the $\bar\gamma_\mu$ matrices.\footnote{In the HV scheme, the Lorentz index of the Dirac matrices $\bar\gamma_\mu$ is restricted to the four space-time coordinates, but the matrices are still infinite dimensional~\cite{Collins:1984xc,Belusca-Maito:2023wah}.} In Ref.~\cite{Naterop:2023dek}, we classified the entire list of evanescent operators that are required in the LEFT up to dimension six at next-to-leading-log (NLL) accuracy. In general, there exist an infinite number of evanescent operators~\cite{Dugan:1990df}. We reproduce the operators relevant for the $\Delta B \neq 0$ sector at NLL in Tab.~\ref{tab:EvanescentFourFermionOperatorsFierzBL} and \ref{tab:EvanescentFourFermionOperatorsBL}.

\input{include/operatorsevanescentHV.tex}

\subsection{Evanescent operators in the NDR scheme}
\label{sec:NDREvanescents}

In the NDR scheme, the prescription for $\gamma_5$ is naively extended to the extra-dimensional sector,
\begin{equation}
    \{ \gamma_5, \gamma_\mu \} = 0 \quad \text{for} \quad \mu = 0,...,D-1 \, ,
\end{equation}
a procedure that in general leads to algebraic inconsistencies~\cite{Jegerlehner:2000dz}, since it is not compatible with both analytic continuation away from $D=4$ and cyclicity of the trace in Dirac space. However, in the present two-loop calculation the only fermionic traces come from vacuum-polarization corrections to the four-fermion $\Delta B \neq 0$ operators. The vacuum polarization does not lead to ill-defined traces, and therefore the NDR scheme can be applied consistently while retaining trace cyclicity. For this reason, we provide the results also in the NDR scheme. 

Similarly to the HV scheme, the NDR scheme leads to two different types of evanescent operators at dimension six. The first type are the Fierz-evanescent operators, which are very similar to those in the HV scheme and obtained from Tab.~\ref{tab:EvanescentFourFermionOperatorsFierzBL} by the substitution $\bar\sigma^{\mu\nu} \mapsto \sigma^{\mu\nu}$, i.e., by undoing the dimensional split. The second type are Chisholm-evanescent operators, which arise due to the non-validity of the Chisholm identity away from $D=4$, preventing the reduction of four-fermion structures with higher numbers of Dirac matrices in each bilinear to lower structures: in NDR, generic four-fermion Dirac structures
\begin{equation}
	\gamma^{\mu_1} \cdots \gamma^{\mu_n} P_{L,R} \otimes \gamma_{\mu_1} \cdots \gamma_{\mu_n} P_{L,R}
\end{equation}
are reduced to structures with a lower number of Dirac matrices, plus an evanescent Dirac structure. Since the evanescent structures can be arbitrarily shifted by $\O(\varepsilon)$ terms, this reduction allows for different evanescent schemes~\cite{Herrlich:1994kh}. In the generic scheme of Ref.~\cite{Dekens:2019ept}, this choice is parametrized by constants $a_\mathrm{ev}$, $b_\mathrm{ev}$, \ldots, where a popular scheme~\cite{Tracas:1982gp,Buras:1989xd} (sometimes called ``Greek projection'') corresponds to $a_\mathrm{ev} = b_\mathrm{ev} = \ldots = 1$.  

In the context of fermion-number-violating interactions, which are most easily treated in the formalism of Ref.~\cite{Denner:1992vza}, we encounter a subtlety, as already observed in Ref.~\cite{Dekens:2019ept}. Since fermion number is violated, there is no canonical direction of fermion-number flow and Feynman diagrams are evaluated by choosing an arbitrary reading direction for fermion lines. If the reading direction agrees with the fermion-number flow, the usual Feynman rules apply. For the part of the diagram where the reading direction is opposite to the fermion-number flow, reversed vertices, propagators, and spinors have to be inserted. Flipping the arbitrary reading direction corresponds to an overall transposition of a Dirac chain, but leads to the same result. As observed in Ref.~\cite{Dekens:2019ept} in the context of Majorana-neutrino operators, the evanescent Dirac structures defined in a generic scheme do not necessarily transform in a simple way under transposition of a Dirac bilinear, which implies that one needs to keep track of the reading direction when identifying evanescent structures. This becomes most transparent if the reduction rules are written in terms of antisymmetrized products of Dirac matrices~\cite{Dugan:1990df}. Defining
\begin{equation}
	\gamma^{[\mu_1}\gamma^{\mu_2} \cdots \gamma^{\mu_n]} =\frac{1}{n!}  \Big( \gamma^{\mu_1}\gamma^{\mu_2} \cdots \gamma^{\mu_n} \; + \; \text{alternating sum over permutations} \Big) \, ,
\end{equation}
we rewrite the reduction rules of Ref.~\cite{Dekens:2019ept} as
\begin{align}
	\label{eq:NDREvanescentsAntisymm}
	\gamma^{[\mu} \gamma^{\nu]} P_L \otimes \gamma_{[\mu} \gamma_{\nu]} P_R &= 2(1+2a_\mathrm{ev}) \varepsilon \, P_L \otimes P_R + \mathcal{E}^{(2)}_{LR} \, ,\nn
	\gamma^{[\mu} \gamma^\nu \gamma^{\rho]} P_L  \otimes \gamma_{[\mu} \gamma_\nu \gamma_{\rho]} P_L &= 2(+3+(3-2b_\mathrm{ev})\varepsilon) \, \gamma^\mu P_L \otimes \gamma_\mu P_L + \mathcal{E}^{(3)}_{LL} \, ,\nn
	\gamma^{[\mu} \gamma^\nu \gamma^{\rho]} P_L  \otimes \gamma_{[\mu} \gamma_\nu \gamma_{\rho]} P_R &= 2(-3+(3+2c_\mathrm{ev})\varepsilon) \, \gamma^\mu P_L \otimes \gamma_\mu P_R + \mathcal{E}^{(3)}_{LR} \, ,\nn
	\gamma^{[\mu} \gamma^\nu \gamma^\rho \gamma^{\sigma]} P_L  \otimes \gamma_{[\mu} \gamma_\nu \gamma_\rho \gamma_{\sigma]} P_L &= 4(6+(11-24 d_\mathrm{ev})\varepsilon) \,P_L \otimes P_L \nn
        &\quad + 4(-3+2 e_\mathrm{ev}) \varepsilon \, \sigma^{\mu\nu} P_L \otimes \sigma_{\mu\nu} P_L + \mathcal{E}^{(4)}_{LL} \, ,\nn
	\gamma^{[\mu} \gamma^\nu \gamma^\rho \gamma^{\sigma]} P_L  \otimes \gamma_{[\mu} \gamma_\nu \gamma_\rho \gamma_{\sigma]} P_R &= 4(-6+(3-16a_\mathrm{ev}+32f_\mathrm{ev})\varepsilon) \, P_L \otimes P_R + \mathcal{E}_{LR}^{(4)} \, ,
\end{align}
where the evanescent structures defined by Eq.~\eqref{eq:NDREvanescentsAntisymm} are related to the ones of Ref.~\cite{Dekens:2019ept} by
\begin{align}
	\E_{LR}^{(2)} &= E_{LR}^{(2),\text{\cite{Dekens:2019ept}}} \, , \nn
	\E_{LL}^{(3)} &= E_{LL}^{(3),\text{\cite{Dekens:2019ept}}} \, , \nn
	\E_{LR}^{(3)} &= E_{LR}^{(3),\text{\cite{Dekens:2019ept}}} \, , \nn
	\E_{LL}^{(4)} &= E_{LL}^{(4),\text{\cite{Dekens:2019ept}}} \, , \nn
	\E_{LR}^{(4)} &= E_{LR}^{(4),\text{\cite{Dekens:2019ept}}} - 4(4-3\varepsilon) E_{LR}^{(2),\text{\cite{Dekens:2019ept}}} + \O(\varepsilon^2) \, .
\end{align}
The antisymmetrized products of Dirac matrices have simple transposition rules:
\begin{align}
	\label{eq:GammaTransposition}
	\left( \gamma^{[\mu_1}\gamma^{\mu_2} \cdots \gamma^{\mu_{2n}]} \right)^T &= (-1)^{n+1} C \gamma^{[\mu_1}\gamma^{\mu_2} \cdots \gamma^{\mu_{2n}]} C \, , \nn
	\left( \gamma^{[\mu_1}\gamma^{\mu_2} \cdots \gamma^{\mu_{2n+1}]} \right)^T &= (-1)^{n} C \gamma^{[\mu_1}\gamma^{\mu_2} \cdots \gamma^{\mu_{2n+1}]} C \, ,
\end{align}
where $C$ is the charge-conjugation matrix. Due to the sign that alternates with the addition of two Dirac matrices, this implies that $\E_{LL}^{(3)}$ and $\E_{LR}^{(3)}$ are trivially related by transposition only for $b_\mathrm{ev} = 3 + c_\mathrm{ev}$~\cite{Dekens:2019ept}. Similarly, in order to combine only terms transforming identically under transposition, one has to choose $a_\mathrm{ev} = -1/2$ and $e_\mathrm{ev} = 3/2$. This is the scheme that we choose in the present calculation, where we only encounter even numbers of Dirac matrices. It simplifies the calculation, as the identification of evanescent terms on a diagram by diagram basis becomes insensitive to the reading direction, which can be assigned differently in each diagram. We note that the scheme corresponding to the ``Greek projection'' does not lead to simple transformation rules of the evanescent structures under transposition. Such a scheme can still be used, but it obscures flavor symmetries of the operator coefficients, discussed in the next section. It also requires one to explicitly keep track of the fermion-chain reading direction, in order not to misidentify evanescent terms. Therefore, for the baryon-number-violating sector of the LEFT, the calculation is done most easily in an evanescent NDR scheme with $a_\mathrm{ev} = -1/2$ and $e_\mathrm{ev} = 3/2$. For similar reasons, in Ref.~\cite{Naterop:2023dek} we defined the evanescent operators in the HV scheme for the $B$-violating sector with antisymmetrized evanescent indices.

We enumerate the Chisholm-evanescent operators in the NDR scheme that are potentially needed at NLL in Tab.~\ref{tab:EvanescentFourFermionOperatorsBLNDR}. The explicit form of the operators is obtained by combining the evanescent Dirac structures of Eq.~\eqref{eq:NDREvanescentsAntisymm} for $a_\mathrm{ev} = -1/2$ and $e_\mathrm{ev} = 3/2$ with the field content of the physical operators in Tab.~\ref{tab:PhysicalOperators}, e.g.,
\begin{equation}
	\label{eq:ExampleNamingConvention}
	\E^{(2),LR}_{\substack{ddd \\ prst}} = \epsilon_{\alpha\beta\gamma} ( d^{\alpha T}_{Lp} C \gamma^{[\mu} \gamma^{\nu]} d^\beta_{Lr}) ( \bar e_{Ls} \gamma_{[\mu} \gamma_{\nu]} d^\gamma_{Rr}) \, ,
\end{equation}
where $p$, $r$, $s$, and $t$ are flavor indices.

In addition to the listed ones, further evanescent operators can potentially be generated in a matching calculation to the LEFT, e.g., Fierz-evanescent operators relating vector--vector four-fermion structures with mixed chiralities to scalar ones. We do not list them, as they have been eliminated as Fierz-redundant operators from the basis of Ref.~\cite{Jenkins:2017jig} and at dimension six they cannot reappear within the renormalization of the LEFT. In case that they are generated in a matching calculation, they should be properly renormalized as usual~\cite{Dugan:1990df,Herrlich:1994kh}.

\input{include/operatorsevanescentNDR.tex}

%% file: include/operatorsphysical.tex

\begin{table}[t]
\capstart
\centering
\scalebox{0.8}{
\begin{minipage}[t]{3cm}
\renewcommand{\arraystretch}{1.5}
\small
\begin{align*}
\begin{array}[t]{c|c}
\multicolumn{2}{c}{\boldsymbol{\Delta B = \Delta L = 1 + \hc}} \\
\hline
\op{udd}{S}{LL} &  \epsilon_{\alpha\beta\gamma}  (u_{Lp}^{\alpha T} C d_{Lr}^{\beta}) (d_{Ls}^{\gamma T} C \nu_{Lt}^{})   \\
\op{duu}{S}{LL} & \epsilon_{\alpha\beta\gamma}  (d_{Lp}^{\alpha T} C u_{Lr}^{\beta}) (u_{Ls}^{\gamma T} C e_{Lt}^{})  \\
\op{uud}{S}{LR} & \epsilon_{\alpha\beta\gamma}  (u_{Lp}^{\alpha T} C u_{Lr}^{\beta}) (d_{Rs}^{\gamma T} C e_{Rt}^{})  \\
\op{duu}{S}{LR} & \epsilon_{\alpha\beta\gamma}  (d_{Lp}^{\alpha T} C u_{Lr}^{\beta}) (u_{Rs}^{\gamma T} C e_{Rt}^{})   \\
\op{uud}{S}{RL} & \epsilon_{\alpha\beta\gamma}  (u_{Rp}^{\alpha T} C u_{Rr}^{\beta}) (d_{Ls}^{\gamma T} C e_{Lt}^{})   \\
\op{duu}{S}{RL} & \epsilon_{\alpha\beta\gamma}  (d_{Rp}^{\alpha T} C u_{Rr}^{\beta}) (u_{Ls}^{\gamma T} C e_{Lt}^{})   \\
\op{dud}{S}{RL} & \epsilon_{\alpha\beta\gamma}  (d_{Rp}^{\alpha T} C u_{Rr}^{\beta}) (d_{Ls}^{\gamma T} C \nu_{Lt}^{})   \\
\op{ddu}{S}{RL} & \epsilon_{\alpha\beta\gamma}  (d_{Rp}^{\alpha T} C d_{Rr}^{\beta}) (u_{Ls}^{\gamma T} C \nu_{Lt}^{})   \\
\op{duu}{S}{RR}  & \epsilon_{\alpha\beta\gamma}  (d_{Rp}^{\alpha T} C u_{Rr}^{\beta}) (u_{Rs}^{\gamma T} C e_{Rt}^{})  \\
\end{array}
\end{align*}
\end{minipage}
%
\begin{minipage}[t]{3cm}
\renewcommand{\arraystretch}{1.5}
\small
\begin{align*}
\begin{array}[t]{c|c}
\multicolumn{2}{c}{\boldsymbol{\Delta B = - \Delta L = 1 + \hc}}  \\
\hline
\op{ddd}{S}{LL} & \epsilon_{\alpha\beta\gamma}  (d_{Lp}^{\alpha T} C d_{Lr}^{\beta}) (\bar e_{Rs}^{} d_{Lt}^\gamma )  \\
\op{udd}{S}{LR}  & \epsilon_{\alpha\beta\gamma}  (u_{Lp}^{\alpha T} C d_{Lr}^{\beta}) (\bar \nu_{Ls}^{} d_{Rt}^\gamma )  \\
\op{ddu}{S}{LR} & \epsilon_{\alpha\beta\gamma}  (d_{Lp}^{\alpha T} C d_{Lr}^{\beta})  (\bar \nu_{Ls}^{} u_{Rt}^\gamma )  \\
\op{ddd}{S}{LR} & \epsilon_{\alpha\beta\gamma}  (d_{Lp}^{\alpha T} C d_{Lr}^{\beta}) (\bar e_{Ls}^{} d_{Rt}^\gamma ) \\
\op{ddd}{S}{RL}  & \epsilon_{\alpha\beta\gamma}  (d_{Rp}^{\alpha T} C d_{Rr}^{\beta}) (\bar e_{Rs}^{} d_{Lt}^\gamma )  \\
\op{udd}{S}{RR}  & \epsilon_{\alpha\beta\gamma}  (u_{Rp}^{\alpha T} C d_{Rr}^{\beta}) (\bar \nu_{Ls}^{} d_{Rt}^\gamma )  \\
\op{ddd}{S}{RR}  & \epsilon_{\alpha\beta\gamma}  (d_{Rp}^{\alpha T} C d_{Rr}^{\beta}) (\bar e_{Ls}^{} d_{Rt}^\gamma )  \\
\end{array}
\end{align*}
\end{minipage}
%
}
\caption{LEFT operators of dimension six that violate baryon number. $p$, $r$, $s$, and $t$ denote flavor indices.}
\label{tab:PhysicalOperators}
\end{table}

%% file: include/operatorsevanescentHV.tex

\begin{table}[t]
    \capstart
    \centering

    \begin{adjustbox}{width=0.75\textwidth,center}
        \begin{minipage}[t]{3cm}
        \renewcommand{\arraystretch}{1.5}
        \small
        \begin{align*}
        \begin{array}[t]{c|c}
        \multicolumn{2}{c}{\boldsymbol{\Delta B = \Delta L = 1 + \hc}}  \\
        \hline
        \EOp{duu}{LL}[(F2)][] & \epsilon_{\alpha\beta\gamma} \Big[ ( d_{Lp}^{\alpha T} C \bar\sigma^{\mu\nu} u^\beta_{Lr}) (u_{Ls}^{\gamma T} C \bar\sigma_{\mu\nu} e_{Lt}) - 4 (d_{Lp}^{\alpha T} C u_{Lr}^\beta) (u_{Ls}^{\gamma T} C e_{Lt}) + 8 (d_{Lp}^{\alpha T} C u_{Ls}^\beta ) (u_{Lr}^{\gamma T} C e_{Lt}) \Big]  \\
        \EOp{duu}{RR}[(F2)][] & \epsilon_{\alpha\beta\gamma} \Big[ ( d_{Rp}^{\alpha T} C \bar\sigma^{\mu\nu} u^\beta_{Rr}) (u_{Rs}^{\gamma T} C \bar\sigma_{\mu\nu} e_{Rt}) - 4 (d_{Rp}^{\alpha T} C u_{Rr}^\beta) (u_{Rs}^{\gamma T} C e_{Rt}) + 8 (d_{Rp}^{\alpha T} C u_{Rs}^\beta ) (u_{Rr}^{\gamma T} C e_{Rt}) \Big]  \\
        \EOp{udd}{LL}[(F2)][] & \epsilon_{\alpha\beta\gamma} \Big[ ( u_{Lp}^{\alpha T} C \bar\sigma^{\mu\nu} d^\beta_{Lr}) (d_{Ls}^{\gamma T} C \bar\sigma_{\mu\nu} \nu_{Lt}) - 4 (u_{Lp}^{\alpha T} C d_{Lr}^\beta) (d_{Ls}^{\gamma T} C \nu_{Lt}) + 8 (u_{Lp}^{\alpha T} C d_{Ls}^\beta) (d_{Lr}^{\gamma T} C \nu_{Lt}) \Big]  \\
        \end{array}
        \end{align*}
        \end{minipage}
        \end{adjustbox}
        
        \begin{adjustbox}{width=0.75\textwidth,center}
        \begin{minipage}[t]{3cm}
        \renewcommand{\arraystretch}{1.5}
        \small
        \begin{align*}
        \begin{array}[t]{c|c}
        \multicolumn{2}{c}{\boldsymbol{\Delta B = - \Delta L = 1 + \hc}}  \\
        \hline
        \EOp{udd}{RR}[(F2)][] & \epsilon_{\alpha\beta\gamma} \Big[ ( u_{Rp}^{\alpha T} C \bar\sigma^{\mu\nu} d^\beta_{Rr}) (\bar \nu_{Ls} \bar\sigma_{\mu\nu} d_{Rt}^\gamma) + 4 (u_{Rp}^{\alpha T} C d_{Rr}^\beta) (\bar\nu_{Ls} d_{Rt}^\gamma) - 8 (u_{Rp}^{\alpha T} C d_{Rt}^\beta ) (\bar\nu_{Ls} d_{Rr}^\gamma) \Big]  \\
        \end{array}
        \end{align*}
        \end{minipage}
        \end{adjustbox}
        
        \caption{Fierz-evanescent LEFT operators that appear in the HV scheme and violate baryon number. Additional Fierz-evanescent operators generally exist (see Ref.~\cite{Naterop:2023dek}) but are not needed as divergent counterterms at NLL. The operators used for our NDR calculation are obtained by replacing $\bar\sigma^{\mu\nu}$ by the $D$-dimensional Dirac tensor $\sigma^{\mu\nu}$.}
        \label{tab:EvanescentFourFermionOperatorsFierzBL}

    \scalebox{0.8}{
    \begin{minipage}[t]{3cm}
    \renewcommand{\arraystretch}{1.5}
    \small
    \begin{align*}
    \begin{array}[t]{c|c}
    \multicolumn{2}{c}{\boldsymbol{\Delta B = \Delta L = 1 + \hc}} \\
    \hline
    \EOp{udd}{LL}[S(T)][] &  \epsilon_{\alpha\beta\gamma}  (u_{Lp}^{\alpha T} C \hat\sigma^{\mu\nu} d_{Lr}^{\beta}) (d_{Ls}^{\gamma T} C \hat\sigma_{\mu\nu} \nu_{Lt}^{})   \\
    \EOp{udd}{LL}[V(1)][] &  \epsilon_{\alpha\beta\gamma}  (u_{Rp}^{\alpha T} C \hat\gamma^\mu \bar\gamma^\nu d_{Lr}^{\beta}) (d_{Rs}^{\gamma T} C \hat\gamma_\mu \bar\gamma_\nu \nu_{Lt}^{})   \\
    \EOp{dud}{LL}[V(1)][] & \epsilon_{\alpha\beta\gamma}  (d_{Rp}^{\alpha T} C \hat\gamma^\mu \bar\gamma^\nu u_{Lr}^{\beta}) (d_{Rs}^{\gamma T} C  \hat\gamma_\mu \bar\gamma_\nu \nu_{Lt}^{})   \\
    \EOp{duu}{LL}[S(T)][] & \epsilon_{\alpha\beta\gamma}  (d_{Lp}^{\alpha T} C \hat\sigma^{\mu\nu} u_{Lr}^{\beta}) (u_{Ls}^{\gamma T} C \hat\sigma_{\mu\nu} e_{Lt}^{})  \\
    \EOp{duu}{LL}[V(1)][] & \epsilon_{\alpha\beta\gamma}  (d_{Rp}^{\alpha T} C \hat\gamma^\mu \bar\gamma^\nu u_{Lr}^{\beta}) (u_{Rs}^{\gamma T} C \hat\gamma_\mu \bar\gamma_\nu e_{Lt}^{})  \\
    \EOp{uud}{LR}[V(1)][] & \epsilon_{\alpha\beta\gamma}  (u_{Rp}^{\alpha T} C \hat\gamma^\mu \bar\gamma^\nu u_{Lr}^{\beta}) (d_{Ls}^{\gamma T} C \hat\gamma_\mu \bar\gamma_\nu e_{Rt}^{})  \\
    \EOp{duu}{LR}[S(T)][] & \epsilon_{\alpha\beta\gamma}  (d_{Lp}^{\alpha T} C \hat\sigma^{\mu\nu} u_{Lr}^{\beta}) (u_{Rs}^{\gamma T} C \hat\sigma_{\mu\nu} e_{Rt}^{})   \\
    \EOp{duu}{LR}[V(1)][] & \epsilon_{\alpha\beta\gamma}  (d_{Rp}^{\alpha T} C \hat\gamma^\mu \bar\gamma^\nu u_{Lr}^{\beta}) (u_{Ls}^{\gamma T} C \hat\gamma_\mu \bar\gamma_\nu e_{Rt}^{})   \\
    \EOp{uud}{RL}[V(1)][] & \epsilon_{\alpha\beta\gamma}  (u_{Lp}^{\alpha T} C \hat\gamma^\mu \bar\gamma^\nu u_{Rr}^{\beta}) (d_{Rs}^{\gamma T} C \hat\gamma_\mu \bar\gamma_\nu e_{Lt}^{})   \\
    \EOp{duu}{RL}[S(T)][] & \epsilon_{\alpha\beta\gamma}  (d_{Rp}^{\alpha T} C \hat\sigma^{\mu\nu} u_{Rr}^{\beta}) (u_{Ls}^{\gamma T} C \hat\sigma_{\mu\nu} e_{Lt}^{})   \\
    \EOp{duu}{RL}[V(1)][] & \epsilon_{\alpha\beta\gamma}  (d_{Lp}^{\alpha T} C \hat\gamma^\mu \bar\gamma^\nu u_{Rr}^{\beta}) (u_{Rs}^{\gamma T} C \hat\gamma_\mu \bar\gamma_\nu e_{Lt}^{})   \\
    \EOp{dud}{RL}[S(T)][] & \epsilon_{\alpha\beta\gamma}  (d_{Rp}^{\alpha T} C \hat\sigma^{\mu\nu} u_{Rr}^{\beta}) (d_{Ls}^{\gamma T} C  \hat\sigma_{\mu\nu} \nu_{Lt}^{})   \\
    \EOp{ddu}{RL}[V(1)][] & \epsilon_{\alpha\beta\gamma}  (d_{Lp}^{\alpha T} C \hat\gamma^\mu \bar\gamma^\nu d_{Rr}^{\beta}) (u_{Rs}^{\gamma T} C \hat\gamma_\mu \bar\gamma_\nu \nu_{Lt}^{})   \\
    \EOp{duu}{RR}[S(T)][]  & \epsilon_{\alpha\beta\gamma}  (d_{Rp}^{\alpha T} C \hat\sigma^{\mu\nu} u_{Rr}^{\beta}) (u_{Rs}^{\gamma T} C \hat\sigma_{\mu\nu} e_{Rt}^{})  \\
    \EOp{duu}{RR}[V(1)][]  & \epsilon_{\alpha\beta\gamma}  (d_{Lp}^{\alpha T} C \hat\gamma^\mu \bar\gamma^\nu u_{Rr}^{\beta}) (u_{Ls}^{\gamma T} C \hat\gamma_\mu \bar\gamma_\nu e_{Rt}^{})  \\
    \end{array}
    \end{align*}
    \end{minipage}
    %
    \begin{minipage}[t]{3cm}
    \renewcommand{\arraystretch}{1.5}
    \small
    \begin{align*}
    \begin{array}[t]{c|c}
    \multicolumn{2}{c}{\boldsymbol{\Delta B = - \Delta L = 1 + \hc}}  \\
    \hline
    \EOp{ddd}{LL}[V(1)][] & \epsilon_{\alpha\beta\gamma}  (d_{Rp}^{\alpha T} C \hat\gamma^\mu \bar\gamma^\nu d_{Lr}^{\beta}) (\bar e_{Ls}^{} \hat\gamma_\mu \bar\gamma_\nu d_{Lt}^\gamma )  \\
    \EOp{udd}{LR}[S(T)][]  & \epsilon_{\alpha\beta\gamma}  (u_{Lp}^{\alpha T} C \hat\sigma^{\mu\nu} d_{Lr}^{\beta}) (\bar \nu_{Ls}^{} \hat\sigma_{\mu\nu} d_{Rt}^\gamma )  \\
    \EOp{ud\nu d}{LL}[V(1)][]  & \epsilon_{\alpha\beta\gamma}  (u_{Rp}^{\alpha T} C \hat\gamma^\mu \bar\gamma^\nu d_{Lr}^{\beta}) (\bar \nu_{Ls}^{} \hat\gamma_\mu \bar\gamma_\nu d_{Lt}^\gamma )  \\
    \EOp{ddu}{LL}[V(1)][] & \epsilon_{\alpha\beta\gamma}  (d_{Rp}^{\alpha T} C \hat\gamma^\mu \bar\gamma^\nu d_{Lr}^{\beta})  (\bar \nu_{Ls}^{} \hat\gamma_\mu \bar\gamma_\nu u_{Lt}^\gamma )  \\
    \EOp{ddd}{LR}[V(1)][] & \epsilon_{\alpha\beta\gamma}  (d_{Rp}^{\alpha T} C \hat\gamma^\mu \bar\gamma^\nu d_{Lr}^{\beta}) (\bar e_{Rs}^{} \hat\gamma_\mu \bar\gamma_\nu d_{Rt}^\gamma )  \\
    \EOp{udd}{RR}[S(T)][]  & \epsilon_{\alpha\beta\gamma}  (u_{Rp}^{\alpha T} C \hat\sigma^{\mu\nu} d_{Rr}^{\beta}) (\bar \nu_{Ls}^{} \hat\sigma_{\mu\nu} d_{Rt}^\gamma )  \\
    \EOp{ud\nu d}{RL}[V(1)][]  & \epsilon_{\alpha\beta\gamma}  (u_{Lp}^{\alpha T} C \hat\gamma^\mu \bar\gamma^\nu d_{Rr}^{\beta}) (\bar \nu_{Ls}^{} \hat\gamma_\mu \bar\gamma_\nu d_{Lt}^\gamma )  \\
    \end{array}
    \end{align*}
    \end{minipage}
    }
    \caption{Evanescent four-fermion LEFT operators other than Fierz-evanescent ones that appear at NLL in the HV scheme and violate baryon number.  }
    \label{tab:EvanescentFourFermionOperatorsBL}
 
\end{table}

%% file: include/operatorsevanescentNDR.tex

\begin{table}[t]
    \capstart
    \centering

       \scalebox{0.8}{
    \begin{minipage}[t]{3cm}
    \renewcommand{\arraystretch}{1.5}
    \small
    \begin{align*}
    \begin{array}[t]{l}
    \multicolumn{1}{c}{\boldsymbol{\Delta B = \Delta L = 1 + \hc}} \\
    \hline
    \EOp{duu}{LR}[(2)][] \; , \;
    \EOp{duu}{RL}[(2)][]  \; , \;
    \EOp{uud}{LR}[(2)][] \; , \;
    \EOp{uud}{RL}[(2)][]  \; , \; \\
    \EOp{duu}{LL}[(4)][] \; , \;
    \EOp{duu}{LR}[(4)][] \; , \;
    \EOp{duu}{RL}[(4)][] \; , \;
    \EOp{duu}{RR}[(4)][] \; , \;
    \EOp{uud}{LR}[(4)][] \; , \;
    \EOp{uud}{RL}[(4)][]   \\
    \hline
    \EOp{dud}{RL}[(2)][] \; , \;
    \EOp{ddu}{RL}[(2)][]  \; , \; \\
    \EOp{udd}{LL}[(4)][] \; , \;
    \EOp{dud}{RL}[(4)][] \; , \;
    \EOp{ddu}{RL}[(4)][]   \\
    \hline
    \end{array}
    \end{align*}
    \end{minipage}
    %
    \quad
    \begin{minipage}[t]{3cm}
    \renewcommand{\arraystretch}{1.5}
    \small
    \begin{align*}
    \begin{array}[t]{l}
    \multicolumn{1}{c}{\boldsymbol{\Delta B = - \Delta L = 1 + \hc}}  \\
    \hline
    \EOp{ddd}{LR}[(2)][] \; , \;
    \EOp{ddd}{RL}[(2)][]  \; , \; \\
    \EOp{ddd}{LL}[(4)][] \; , \;
    \EOp{ddd}{LR}[(4)][] \; , \;
    \EOp{ddd}{RL}[(4)][] \; , \;
    \EOp{ddd}{RR}[(4)][]  \\
    \hline
    \EOp{udd}{LR}[(2)][] \; , \;
    \EOp{ddu}{LR}[(2)][] \; , \;  \\
    \EOp{udd}{LR}[(4)][] \; , \;
    \EOp{ddu}{LR}[(4)][] \; , \;
    \EOp{udd}{RR}[(4)][]  \\
    \hline
    \end{array}
    \end{align*}
    \end{minipage}
    }
    \caption{List of evanescent four-fermion LEFT operators other than Fierz-evanescent ones that potentially appear at NLL in the NDR scheme and violate baryon number. The operators are defined by combining the evanescent Dirac structures of Eq.~\eqref{eq:NDREvanescentsAntisymm} for $a_\mathrm{ev} = -1/2$ and $e_\mathrm{ev} = 3/2$ with the field content of the physical operators in Tab.~\ref{tab:PhysicalOperators}, see Eq.~\eqref{eq:ExampleNamingConvention} for an example.}
    \label{tab:EvanescentFourFermionOperatorsBLNDR}
 
\end{table}

%% file: sections/Flavor.tex

\section{Flavor structure and symmetries}
\label{sec:Flavor}

In the LEFT Lagrangian, operators and Wilson coefficients carry flavor indices. E.g., in the case of four-fermion operators, the short-hand notation of Eq.~\eqref{eq:LEFTLagrangian} stands for
\begin{equation}
	\label{eq:SumWCtimesOps}
	\sum_i L_i \O_i = \sum_{p,r,s,t} L_{prst} \O_{prst} \, ,
\end{equation}
with flavor indices $p$, $r$, $s$, and $t$. Many of the LEFT operators exhibit flavor symmetries, e.g., for operators consisting of two identical bilinears, we have $\O_{prst} = \O_{stpr}$. This seemingly leads to redundancies in the operator basis, which however are removed by constraints on the Wilson coefficients. We follow the convention of Refs.~\cite{Jenkins:2017dyc,Jenkins:2017jig,Dekens:2019ept,Naterop:2023dek} and require the Wilson coefficients to fulfill the same symmetries as the operators.

For the $B$-violating operators, the symmetries of the physical operators in four space-time dimensions are
\begin{align}
	\label{eq:FlavorSymmetries}
	\Op{uud}{XY}[S][prst] &= -\Op{uud}{XY}[S][rpst] \, , \quad XY \in \{ LR, RL \} \, , \nn
	\Op{ddu}{XY}[S][prst] &= -\Op{ddu}{XY}[S][rpst] \, , \quad XY \in \{ LR, RL \} \, , \nn
	\Op{ddd}{XY}[S][prst] &= -\Op{ddd}{XY}[S][rpst] \, , \quad XY \in \{ LL, LR, RL, RR \} \, , \nn
	\Op{ddd}{XX}[S][prst] &= \Op{ddd}{XX}[S][trsp] - \Op{ddd}{XX}[S][tpsr] \, , \quad XX \in \{ LL, RR \} \, .
\end{align}
The first three equations hold in general and directly translate into symmetries of the associated Wilson coefficients,
\begin{align}
	\label{eq:FlavorSymmetriesCoeffs}
	\lwc{uud}{XY}[S][prst] &= -\lwc{uud}{XY}[S][rpst] \, , \quad XY \in \{ LR, RL \} \, , \nn
	\lwc{ddu}{XY}[S][prst] &= -\lwc{ddu}{XY}[S][rpst] \, , \quad XY \in \{ LR, RL \} \, , \nn
	\lwc{ddd}{XY}[S][prst] &= -\lwc{ddd}{XY}[S][rpst] \, , \quad XY \in \{ LL, LR, RL, RR \} \, .
\end{align}
Not imposing these constraints (or equivalent ones~\cite{Aebischer:2025hsx}) would not change the Lagrangian, since a contribution to the Wilson coefficients that is symmetric in $p\leftrightarrow r$ drops out in the sum~\eqref{eq:SumWCtimesOps}, but this would indeed introduce redundancies in the set of parameters. Therefore, we impose Eq.~\eqref{eq:FlavorSymmetriesCoeffs} and also use these relations to simplify our results.

The last equation in Eq.~\eqref{eq:FlavorSymmetries} does not hold away from $D=4$, but gives rise to the evanescent operators~\cite{Naterop:2023dek}
\begin{align}
	\label{eq:FierzEvanddd}
	\EOp{ddd}{LL}[(F)][prst] &= \epsilon_{\alpha\beta\gamma} \Big[ ( d_{Lp}^{\alpha T} C d^\beta_{Lr}) (\bar e_{Rs} d_{Lt}^\gamma ) - (d_{Lt}^{\alpha T} C d_{Lr}^\beta) (\bar e_{Rs} d_{Lp}^\gamma ) + (d_{Lt}^{\alpha T} C d_{Lp}^\beta) (\bar e_{Rs} d_{Lr}^\gamma) \Big] \, , \nn
	\EOp{ddd}{RR}[(F)][prst] &= \epsilon_{\alpha\beta\gamma} \Big[ ( d_{Rp}^{\alpha T} C d^\beta_{Rr}) (\bar e_{Ls} d_{Rt}^\gamma ) - (d_{Rt}^{\alpha T} C d_{Rr}^\beta) (\bar e_{Ls} d_{Rp}^\gamma ) + (d_{Rt}^{\alpha T} C d_{Rp}^\beta) (\bar e_{Ls} d_{Rr}^\gamma) \Big]  \, .
\end{align}
We adopt the scheme that the Wilson coefficient nevertheless satisfies the flavor symmetry that holds for the operator in $D=4$ dimensions. A contribution violating that symmetry is reabsorbed into the coefficient of the evanescent operators. Explicitly, consider the situation of a coefficient $L_{prst}$ without manifest flavor symmetry. We construct a linear combination $L_{prst}^\text{sym}$ of the flavor permutations of $L_{prst}$ and require $L_{prst}^\text{sym} = -L_{rpst}^\text{sym}$, $L_{prst}^\text{sym} = L_{trsp}^\text{sym} - L_{tpsr}^\text{sym}$, as well as
\begin{equation}
	L_{prst} \O_{prst} = L_{prst}^\text{sym} \O_{prst} + \text{evanescent} \, ,
\end{equation}
which results in
\begin{equation}
	\label{eq:Symmetrization}
	L_{prst}^\text{sym} = \frac{1}{6} (2L_{prst} - 2L_{rpst} + L_{ptsr} - L_{rtsp} + L_{trsp} - L_{tpsr}) \, .
\end{equation}
In particular, with $L_{prst} = L_{prst}^\text{sym} + L_{prst}^\text{asym}$, one finds
\begin{equation}
	\label{eq:AsymmetricFierzEvanShift}
	L_{prst}^\text{asym} \O_{prst} = \frac{1}{3} L_{prst} \E_{prst} \, , \quad \E_{prst} = \O_{prst} - \O_{trsp} + \O_{tpsr} \, ,
\end{equation}
i.e., any contribution not satisfying the flavor symmetries can be shifted into the evanescent coefficient.\footnote{
In Ref.~\cite{Naterop:2023dek}, we erroneously did not apply the Fierz symmetry to the finite counterterms to $\lwc{ddd}{LL}[S]$ and $\lwc{ddd}{RR}[S]$. According to Eq.~\eqref{eq:AsymmetricFierzEvanShift}, this corresponds to a finite shift in the counterterms to the evanescent operators~\eqref{eq:FierzEvanddd}. This is a scheme change compared to what is described in Ref.~\cite{Naterop:2023dek}, i.e., a scheme without any finite counterterms to evanescent operators. For the present computation of the two-loop RGEs, we fix this issue by applying the symmetrization~\eqref{eq:Symmetrization} to the finite counterterms of the two affected coefficients.}

Symmetries restrict the possible form of the RGEs. Chiral symmetry is preserved by the regulator in the NDR scheme. In the HV scheme, it is broken by the regulator but restored by finite counterterms in the scheme defined in Ref.~\cite{Naterop:2023dek}. Therefore, chiral symmetry implies in both schemes that the RGEs do not mix the 16 operator classes of the $B$-violating sector. In our choice of basis, the flavor symmetries further restrict the RGEs. In particular, from the discussion above it follows that the RGEs for $\lwc{ddd}{LL}[S][prst]$ and $\lwc{ddd}{RR}[S][prst]$ can only depend on $\lwc{ddd}{LL}[S][prst]$ and $\lwc{ddd}{RR}[S][prst]$, respectively. For several other classes, the flavor symmetries are simpler and directly imply that the RGEs have to be diagonal, with the exception of the four classes $\lwc{udd}{LL/RR}[S]$, $\lwc{duu}{LL/RR}[S]$, which do not contain any flavor symmetry under exchange of the two down-type or up-type quarks, respectively. Therefore, the RGEs for these four operator classes mix the two crossed structures. This is the flavor structure of the one-loop RGEs computed in Ref.~\cite{Jenkins:2017dyc}. The symmetry arguments are general, hence with our choice of operator basis, the RGEs obey the same flavor structure at any loop order.

In Ref.~\cite{Aebischer:2025hsx}, a different basis is used where the Wilson coefficients do not obey the same flavor symmetries as the operators and redundancies are instead avoided by explicitly restricting the ranges of flavor indices.

The flavor symmetries have further interesting consequences for the evanescent sector. Due to Eq.~\eqref{eq:GammaTransposition}, in the NDR scheme one finds flavor symmetries in the evanescent sector, such as
\begin{align}
	\label{eq:FlavorSymmetriesEvan}
	\EOp{uud}{XY}[(2)][prst] &= \EOp{uud}{XY}[(2)][rpst] \, , \quad XY \in \{ LR, RL \} \, , \nn
	\EOp{ddu}{XY}[(2)][prst] &= \EOp{ddu}{XY}[(2)][rpst] \, , \quad XY \in \{ LR, RL \} \, ,
\end{align}
which translate into analogous symmetries for their coefficients $K_{uud}^{(2),XY}$ and $K_{ddu}^{(2),XY}$. The opposite sign compared to the flavor symmetry of the physical operators immediately implies that these evanescent operators are not generated from insertions of physical operators and for these flavor structures one expects to find evanescent operators only starting with four Dirac matrices, which cannot be generated at one loop. Analogous arguments apply to several Fierz-evanescent operators listed in Ref.~\cite{Naterop:2023dek}.

%% file: sections/NDR1Loop.tex

\section{One-loop renormalization in the NDR scheme}
\label{sec:OneLoopNDR}

The divergent one-loop counterterms to physical operators are scheme independent and have been computed to dimension six in Ref.~\cite{Jenkins:2017dyc} (and independently confirmed in Ref.~\cite{Naterop:2023dek}). The complete one-loop counterterms in the chirally symmetric HV scheme have been derived in Ref.~\cite{Naterop:2023dek}. In the following, we provide the counterterms related to evanescent operators in the NDR scheme for the dimension-six $B$-violating sector.

With the insertion of physical operators, the following non-vanishing divergent counterterms to evanescent operators are generated, consistent with the flavor-symmetry arguments of the previous section:\footnote{For the notation of counterterms, see Ref.~\cite{Naterop:2023dek}. We further use the abbreviation $\{X\}_1 = X / (16\pi^2)$.}
\begin{align}
	\label{eq:NDR1LEvanescentDivergences}
	\delta_\mathrm{div}\Bigl( \kwc{duu}{LR}[(2)][prst] \Bigr) &= \left\{ \frac{e^2 \left(\q_d-\q_u\right) \left(\q_u-\q_e\right)}{4 \varepsilon } \right\}_1 \lwc{duu}{LR}[S][prst]  \,, \nn
	\delta_\mathrm{div}\Bigl( \kwc{duu}{RL}[(2)][prst] \Bigr) &= \left\{ \frac{e^2 \left(\q_d-\q_u\right) \left(\q_u-\q_e\right)}{4 \varepsilon } \right\}_1 \lwc{duu}{RL}[S][prst] \,, \nn
	\delta_\mathrm{div}\Bigl( \kwc{udd}{LR}[(2)][prst] \Bigr) &= \left\{ \frac{e^2 \q_d \left(\q_d-\q_u\right)}{4 \varepsilon } \right\}_1 \lwc{udd}{LR}[S][prst] \,, \nn
	\delta_\mathrm{div}\Bigl( \kwc{dud}{RL}[(2)][prst] \Bigr) &= \left\{ \frac{e^2 \q_d \left(\q_d-\q_u\right)}{4 \varepsilon } \right\}_1 \lwc{dud}{RL}[S][prst] \,, \nn
	\delta_\mathrm{div}\Bigl( \kwc{duu}{LL}[(F2)][prst] \Bigr) &= \left\{ \frac{e^2 \left(\q_d-\q_u\right) \left(\q_e-\q_u\right)}{4 \varepsilon } \right\}_1 \lwc{duu}{LL}[S][prst] \,, \nn
	\delta_\mathrm{div}\Bigl( \kwc{duu}{RR}[(F2)][prst] \Bigr) &= \left\{ \frac{e^2 \left(\q_d-\q_u\right) \left(\q_e-\q_u\right)}{4 \varepsilon } \right\}_1 \lwc{duu}{RR}[S][prst] \,, \nn
	\delta_\mathrm{div}\Bigl( \kwc{udd}{LL}[(F2)][prst] \Bigr) &= \left\{ \frac{e^2 \q_d \left(\q_d-\q_u\right)}{4 \varepsilon } \right\}_1 \lwc{udd}{LL}[S][prst] \,, \nn
	\delta_\mathrm{div}\Bigl( \kwc{udd}{RR}[(F2)][prst] \Bigr) &= \left\{ \frac{e^2 \q_d \left(\q_u-\q_d\right)}{4 \varepsilon } \right\}_1 \lwc{udd}{RR}[S][prst] \, .
\end{align}
No other evanescent divergences are generated. Naively, one also finds divergent contributions to the Fierz-evanescent operators in Eq.~\eqref{eq:FierzEvanddd},
\begin{align}
	\delta_\mathrm{div}\Bigl( \kwc{ddd}{LL}[F][prst] \Bigr) &= \left\{ \frac{3 e^2 \q_d \left(\q_d-\q_e\right) - 2 g^2}{18 \varepsilon } \right\}_1 \left( \lwc{ddd}{LL}[S][prst] - \lwc{ddd}{LL}[S][rpst] + \lwc{ddd}{LL}[S][rtsp] - \lwc{ddd}{LL}[S][ptsr] + \lwc{ddd}{LL}[S][tpsr] - \lwc{ddd}{LL}[S][trsp] \right)  \,, \nn
	\delta_\mathrm{div}\Bigl( \kwc{ddd}{RR}[F][prst] \Bigr) &= \left\{ \frac{3 e^2 \q_d \left(\q_d-\q_e\right) - 2 g^2}{18 \varepsilon } \right\}_1 \left( \lwc{ddd}{RR}[S][prst] - \lwc{ddd}{RR}[S][rpst] + \lwc{ddd}{RR}[S][rtsp] - \lwc{ddd}{RR}[S][ptsr] + \lwc{ddd}{RR}[S][tpsr] - \lwc{ddd}{RR}[S][trsp] \right)  \,,
\end{align}
however, due to the flavor symmetries of the Wilson coefficients in our basis, these expressions vanish. Therefore, in the NDR scheme evanescent counterterms only appear due to QED corrections, whereas the QCD corrections do not lead to any evanescent contributions.

Insertions of evanescent operators into one-loop diagrams can generate finite physical effects, which are compensated by finite physical counterterms in order to avoid a mixing of the evanescent into the physical sector~\cite{Dugan:1990df,Herrlich:1994kh}. Here, we only list the counterterms to insertions of the evanescent operators~\eqref{eq:NDR1LEvanescentDivergences}, which are generated in the LEFT renormalization at one loop. In a matching calculation, one can potentially generate further evanescent operators, which in this case should also be properly renormalized. We find the finite counterterms
\begin{align}
	\label{eq:NDR1LFiniteCTEvan}
	\delta_\mathrm{fin}^\mathrm{ev}\Bigl(\lwc{duu}{LR}[S][prst]\Bigr) &= \left\{ 4 e^2 \left(1 + 8 f_\mathrm{ev}\right) \left(\q_d-\q_u\right) \left(\q_u-\q_e\right) \right\}_1 \kwc{duu}{LR}[(2)][prst] \,, \nn*
	\delta_\mathrm{fin}^\mathrm{ev}\Bigl(\lwc{duu}{RL}[S][prst]\Bigr) &= \left\{ 4 e^2 \left(1 + 8 f_\mathrm{ev}\right) \left(\q_d-\q_u\right) \left(\q_u-\q_e\right) \right\}_1 \kwc{duu}{RL}[(2)][prst] \,, \nn*
	\delta_\mathrm{fin}^\mathrm{ev}\Bigl(\lwc{udd}{LR}[S][prst]\Bigr) &= \left\{ 4 e^2 \left(1 + 8 f_\mathrm{ev}\right) \q_d \left(\q_d-\q_u\right) \right\}_1 \kwc{udd}{LR}[(2)][prst] \,, \nn*
	\delta_\mathrm{fin}^\mathrm{ev}\Bigl(\lwc{dud}{RL}[S][prst]\Bigr) &= \left\{ 4 e^2 \left(1 + 8 f_\mathrm{ev}\right) \q_d \left(\q_d-\q_u\right) \right\}_1 \kwc{dud}{RL}[(2)][prst] \,, \nn
	\delta_\mathrm{fin}^\mathrm{ev}\Bigl(\lwc{duu}{LL}[S][prst]\Bigr) &= \left\{ 4 e^2 \left(\q_d-\q_u\right) \left(\q_u-\q_e\right) \right\}_1 \left(\kwc{duu}{LL}[(F2)][prst] \left(1 + 6 d_\mathrm{ev}\right)-4 \kwc{duu}{LL}[(F2)][psrt] \right) \,, \nn*
	\delta_\mathrm{fin}^\mathrm{ev}\Bigl(\lwc{duu}{RR}[S][prst]\Bigr) &= \left\{ 4 e^2 \left(\q_d-\q_u\right) \left(\q_u-\q_e\right) \right\}_1 \left(\kwc{duu}{RR}[(F2)][prst] \left(1 + 6 d_\mathrm{ev}\right)-4 \kwc{duu}{RR}[(F2)][psrt] \right) \,, \nn*
	\delta_\mathrm{fin}^\mathrm{ev}\Bigl(\lwc{udd}{LL}[S][prst]\Bigr) &= \left\{ 4 e^2 \q_d \left(\q_u-\q_d\right) \right\}_1 \left( \kwc{udd}{LL}[(F2)][prst] \left(1 + 6 d_\mathrm{ev}\right) - 4 \kwc{udd}{LL}[(F2)][psrt] \right) \,, \nn*
	\delta_\mathrm{fin}^\mathrm{ev}\Bigl(\lwc{udd}{RR}[S][prst]\Bigr) &= \left\{ 4 e^2 \q_d \left(\q_d-\q_u\right) \right\}_1 \left( \kwc{udd}{RR}[(F2)][prst] \left(1 + 6 d_\mathrm{ev}\right) - 4 \kwc{udd}{RR}[(F2)][psrt] \right) \, .
\end{align}
These finite renormalizations affect the RGEs at the two-loop level.

%% file: sections/Computation.tex

\section{Computation of the two-loop anomalous dimensions}
\label{sec:Computation}

For the computation of the two-loop anomalous dimensions, we make use of a semi-automated tool chain, designed to obtain the complete two-loop RGEs of the LEFT in the HV scheme. Many details have been described in a previous paper, where we presented the complete two-loop LEFT RGEs in the HV scheme at dimension five in the power counting~\cite{Naterop:2024ydo}. Here, we employ the local $\bar R$-operation for the infrared rearrangement, combined with the introduction of an auxiliary dummy mass as an IR regulator~\cite{Chetyrkin:1997fm}. The Feynman diagrams are generated with \texttt{QGRAF}~\cite{Nogueira:1991ex}. For the application of Feynman rules, the simplifications of color and Dirac algebra, and the IR rearrangement, we use our own routines written in \texttt{FORM}~\cite{Vermaseren:2000nd,Ruijl:2017dtg} and \texttt{Symbolica}. We rely on two independent implementations, which enables a cross check of the entire calculation. The computation is performed with generic QED and QCD gauge parameters. The cancellation of gauge-parameter dependence is another strong check on the results. In the $B$-violating sector, it requires to use $N_c = 3$ and electric charge conservation, since otherwise the dimension-six operators would not be gauge invariant.

In the presence of finite renormalizations, the two-loop RGEs are obtained from the master formula (denoting all physical parameters collectively by $L_i$ and using the short-hand notation of Eq.~\eqref{eq:RGENotation})~\cite{Naterop:2023dek}
\begin{equation}
	\label{eq:RGEMasterFormula}
	\left[ \, \dot L_i \, \right]_2 = 
		4 L_i^{(2,1)} - \sum_j 2 L_j^{(1,0)} \frac{\p L_i^{(1,1)}}{\p L_j} - \sum_j 2 L_j^{(1,1)} \frac{\p L_i^{(1,0)}}{\p L_j} - \sum_j  2 K_j^{(1,1)} \frac{\p L_i^{(1,0)}}{\p K_j} \, ,
\end{equation}
where $X_i^{(l,n)}$, $X\in\{L,K\}$, denote the coefficients of the $1/\varepsilon^n$ poles of the $l$-loop counterterms defined by
\begin{equation}
	X_i^\mathrm{ct} = \sum_{l=1}^\infty \sum_{n=0}^l \frac{1}{\varepsilon^n} \frac{1}{(16\pi^2)^l} X_i^{(l,n)}(\{L_j^r(\mu)\},\{K_k^r(\mu)\}) \, .
\end{equation}
In the HV scheme, all four terms in the master formula are relevant due to finite renormalizations that restore chiral symmetry and compensate evanescent-operator insertions. In the NDR scheme, only the first and last terms in Eq.~\eqref{eq:RGEMasterFormula} contribute, since the only finite renormalizations are the ones compensating evanescent insertions, given in Eq.~\eqref{eq:NDR1LFiniteCTEvan}.

%% file: sections/Results.tex

\section{Results}
\label{sec:Results}

\subsection{Structure of the RGEs}

We provide our explicit results for the RGEs of the dimension-six $B$-violating sector of the LEFT in App.~\ref{sec:RGEs}, both in the chirally symmetric HV scheme and in the NDR scheme. As discussed in Sect.~\ref{sec:Flavor}, our choice of operator basis guarantees that the RGEs at any loop order have the same flavor-mixing structure as the one-loop RGEs~\cite{Jenkins:2017dyc}. Our results at two loops indeed exhibit the same renormalization-group-invariant block structure as the one-loop RGEs. We provide our results for a generic number of flavors $n_f$. The dependence on the number of flavors arises from fermionic loops in vacuum-polarization insertions. Therefore, we find it convenient to write the dependence on $n_f$ in terms of the coefficients of the one-loop QED and QCD $\beta$-functions (for $N_c = 3$),
\begin{align}
	\label{eq:OneLoopBetaFunctions}
	b^e_{0,0} &= -\frac{4}{3} \left( n_e \q_e^2 + 3 (n_u \q_u^2 + n_d \q_d^2 ) \right) \, , \nn
	b^g_{0,0} &= 11 - \frac{2}{3} ( n_u + n_d ) \, .
\end{align}
In our results, we keep the fermion charges generic, using only overall charge conservation in order to obtain gauge-parameter-independent results, as mentioned above.

In the HV scheme, it is important to note that the finite renormalizations are crucial to obtain a result respecting chiral symmetry (in the spurion sense), i.e., the RGEs preserve the chiral structure. If one were to apply minimal subtraction in the HV scheme, complicated RGEs would be obtained that would mix operators with different chiralities. These effects would eventually cancel between RGEs, matching, and matrix-element calculations, and hence they would not affect relations between observables. However, it is clearly preferable to work in a scheme where these spurious symmetry-breaking terms are manifestly absent in intermediate steps of the calculation. The cancellation of symmetry-breaking terms happens among all the terms in the master formula~\eqref{eq:RGEMasterFormula}. Therefore, the symmetry of the two-loop RGEs is a strong check on the finite counterterms obtained in Ref.~\cite{Naterop:2023dek}.

In NDR, we find no one-loop evanescent contributions in pure QCD\footnote{We thank S.~Banik for pointing this out to us.} and in both NDR and HV schemes the pure QCD contributions to the two-loop RGEs are diagonal in the operator basis.

\subsection{Scheme dependences}

While exhibiting the same flavor structure, the results in the HV and NDR schemes are different, which is expected: while the divergent one-loop counterterms and hence the one-loop RGEs are scheme independent, this is no longer true at two loops. The finite one-loop renormalizations of Ref.~\cite{Naterop:2023dek} were chosen as the minimal ones that restore chiral spurion symmetry. Interestingly, the difference between the two schemes does not affect the mixed QCD and QED contributions of $\O(g^2 e^2)$, but only the $\O(g^4)$ and $\O(e^4)$ terms. We find that for all the $B$-violating operators it can be written as
\begin{equation}
	\label{eq:NDRvsHV}
	\left[ \dlwc{}{}[][prst] \right]_2^\mathrm{NDR} - \left[ \dlwc{}{}[][prst] \right]_2^\mathrm{HV} = \left( 8 e^4 b^e_{0,0} (\q_p \q_r + \q_s \q_t) - \frac{16}{3} g^4 b^g_{0,0} \right) \lwc{}{}[][prst] \,,
\end{equation}
where $\q_i$ denote the electric charges of all incoming particles, so that $\q_p + \q_r + \q_s + \q_t = 0$. E.g., in the case of the operator $\Op{LL}{ddd}[S][prst]$, we denote $\q_p = \q_r = \q_t = \q_d = -1/3$ and $\q_s = -\q_e = 1$.

In principle, one could perform additional finite one-loop renormalizations that would bring the two-loop RGEs in the HV and NDR schemes into complete agreement~\cite{Ciuchini:1993ks,Ciuchini:1993fk,DiNoi:2023ygk,OlgosoRuiz:2024dzq,DiNoi:2024ajj,DiNoi:2025uan}. These finite renormalizations need to be chirally invariant in the spurion sense, since the symmetry-breaking renormalizations are fixed by the condition that chiral symmetry be restored. In order to link the schemes, one needs to consider both the differences in the evanescent sector and the symmetry-breaking effects. As an example, we consider the $dded$ Green's function. The NDR scheme does not involve evanescent terms in this sector, while in the HV scheme we do encounter evanescent counterterms~\cite{Naterop:2023dek}. However, these counterterms vanish in the limit of purely vector-like (non-chiral) interactions, i.e., for $\lwc{ddd}{LL}[S] = \lwc{ddd}{RL}[S] = \lwc{ddd}{LR}[S] = \lwc{ddd}{RR}[S]$, which makes the discussion of this sector particularly simple. The finite symmetry-restoring counterterms are given by~\cite{Naterop:2023dek}
\begin{align}
	\label{eq:dddFiniteCTs}
	\delta_\mathrm{fin}^\chi\Bigl( \lwc{ddd}{LL}[S][prst] \Bigr) &= \left\{ \frac{8}{3} g^2 - 4 e^2 \q_d^2 \right\}_1 \lwc{ddd}{RL,\mathrm{sym}}[S][prst] + \Bigl\{ 4 e^2 \q_d \q_e \Bigr\}_1 \lwc{ddd}{LR,\mathrm{sym}}[S][prst] \, , \nn
	\delta_\mathrm{fin}^\chi\Bigl( \lwc{ddd}{RL}[S][prst] \Bigr) &= \left\{ \frac{8}{3} g^2 - 4 e^2 \q_d^2 \right\}_1 \lwc{ddd}{LL}[S][prst] + \Bigl\{ 4 e^2 \q_d \q_e \Bigr\}_1 \lwc{ddd}{RR}[S][prst] \, , \nn
	\delta_\mathrm{fin}^\chi\Bigl( \lwc{ddd}{LR}[S][prst] \Bigr) &= \left\{ \frac{8}{3} g^2 - 4 e^2 \q_d^2 \right\}_1 \lwc{ddd}{RR}[S][prst] + \Bigl\{ 4 e^2 \q_d \q_e \Bigr\}_1 \lwc{ddd}{LL}[S][prst] \, , \nn
	\delta_\mathrm{fin}^\chi\Bigl( \lwc{ddd}{RR}[S][prst] \Bigr) &= \left\{ \frac{8}{3} g^2 - 4 e^2 \q_d^2 \right\}_1 \lwc{ddd}{LR,\mathrm{sym}}[S][prst] + \Bigl\{ 4 e^2 \q_d \q_e \Bigr\}_1 \lwc{ddd}{RL,\mathrm{sym}}[S][prst] \, ,
\end{align}
using the notation for the symmetrization of Eq.~\eqref{eq:Symmetrization}. These are the minimal symmetry-restoring counterterms in our scheme of Ref.~\cite{Naterop:2023dek}, consisting of pure chirality-violating contributions. Without violating chiral spurion symmetry, there is the freedom to apply additional finite symmetry-conserving renormalizations: a different scheme choice would be to impose the absence of finite counterterms in the limit of vector-like interactions, resulting in the modified finite renormalizations
\begin{align}
	\widetilde\delta_\mathrm{fin}^\chi\Bigl( \lwc{ddd}{LL}[S][prst] \Bigr) &= \left\{ \frac{8}{3} g^2 - 4 e^2 \q_d^2 \right\}_1 \left( \lwc{ddd}{RL,\mathrm{sym}}[S][prst] - \lwc{ddd}{LL}[S][prst] \right) + \Bigl\{ 4 e^2 \q_d \q_e \Bigr\}_1 \left( \lwc{ddd}{LR,\mathrm{sym}}[S][prst] - \lwc{ddd}{LL}[S][prst] \right) \, , \nn
	\widetilde\delta_\mathrm{fin}^\chi\Bigl( \lwc{ddd}{RL}[S][prst] \Bigr) &= \left\{ \frac{8}{3} g^2 - 4 e^2 \q_d^2 \right\}_1 \left( \lwc{ddd}{LL}[S][prst] - \lwc{ddd}{RL}[S][prst] \right) + \Bigl\{ 4 e^2 \q_d \q_e \Bigr\}_1 \left( \lwc{ddd}{RR}[S][prst] - \lwc{ddd}{RL}[S][prst] \right) \, , \nn
	\widetilde\delta_\mathrm{fin}^\chi\Bigl( \lwc{ddd}{LR}[S][prst] \Bigr) &= \left\{ \frac{8}{3} g^2 - 4 e^2 \q_d^2 \right\}_1 \left( \lwc{ddd}{RR}[S][prst] - \lwc{ddd}{LR}[S][prst] \right) + \Bigl\{ 4 e^2 \q_d \q_e \Bigr\}_1 \left( \lwc{ddd}{LL}[S][prst] - \lwc{ddd}{LR}[S][prst] \right) \, , \\
	\widetilde\delta_\mathrm{fin}^\chi\Bigl( \lwc{ddd}{RR}[S][prst] \Bigr) &= \left\{ \frac{8}{3} g^2 - 4 e^2 \q_d^2 \right\}_1 \left( \lwc{ddd}{LR,\mathrm{sym}}[S][prst] - \lwc{ddd}{RR}[S][prst] \right) + \Bigl\{ 4 e^2 \q_d \q_e \Bigr\}_1 \left( \lwc{ddd}{RL,\mathrm{sym}}[S][prst] - \lwc{ddd}{RR}[S][prst] \right) \, , \nonumber
\end{align}
which differ from the counterterms in Eq.~\eqref{eq:dddFiniteCTs} only by symmetry-conserving contributions. Using these modified counterterms changes the first three terms in the RGE master formula~\eqref{eq:RGEMasterFormula}. We checked that these modifications correspond to the differences in Eq.~\eqref{eq:NDRvsHV} and hence bring the HV results into exact agreement with the NDR results. This is easily understood, as in the limit of vector-like interactions no $\gamma_5$ appears and given the absence of evanescent operators the HV and NDR schemes are algebraically identical in this limit. Chiral spurion symmetry, which is preserved in both schemes, implies that the RGEs for the chiral operators have to agree as well. This establishes the scheme translation between our HV scheme and the NDR scheme for the $dded$ sector. Similar discussions apply to the $udd\nu$ and $uude$ Green's functions, which however are also affected by differences in the evanescent schemes.

Within the NDR scheme, we find that the dependences on the evanescent scheme parametrized in terms of generic constants $d_\mathrm{ev}$ and $f_\mathrm{ev}$ drop out of the two-loop RGEs through a cancellation between the first and last terms in Eq.~\eqref{eq:RGEMasterFormula}. As explained in Sect.~\ref{sec:NDREvanescents}, we perform the calculation with $a_\mathrm{ev} = -1/2$ and $e_\mathrm{ev} = 3/2$, in order to preserve flavor symmetries and allow for an implementation independent of the fermion-line reading direction. For some coefficients, a dependence on the scheme parameter $a_\mathrm{ev}$ is expected, as such a scheme dependence is indeed observed in the one-loop matching at the electroweak scale~\cite{Dekens:2019ept}. Therefore, we reinstate the dependence on $a_\mathrm{ev}$ a posteriori as follows. Generic $a_\mathrm{ev}$-dependent evanescent operators are related to our basis, e.g., by
\begin{equation}
	\EOp{duu}{LR,a_\mathrm{ev}}[(2)][prst] = \EOp{duu}{LR}[(2)][prst] - 2(1+2a_\mathrm{ev}) \varepsilon \, \Op{duu}{LR}[S][prst] \, .
\end{equation}
Instead of doing the calculation in the basis with generic scheme parameter $a_\mathrm{ev}$, we can use our basis with the coefficient of the physical operator shifted as
\begin{equation}
	\lwc{duu}{LR}[S][prst] = \lwc{duu}{LR,a_\mathrm{ev}}[S][prst] - 2(1+2a_\mathrm{ev}) \varepsilon \, \kwc{duu}{LR,a_\mathrm{ev}}[(2)][prst] \, , \qquad \kwc{duu}{LR}[(2)][prst] = \kwc{duu}{LR,a_\mathrm{ev}}[(2)][prst] \, .
\end{equation}
In particular, with Eq.~\eqref{eq:NDR1LEvanescentDivergences} one obtains the finite one-loop renormalizations
\begin{align}
	\delta_\mathrm{fin}^{a_\mathrm{ev}}\Bigl( \lwc{duu}{LR}[S][prst] \Bigr) &= \left\{ (1+2a_\mathrm{ev}) \frac{e^2 \left(\q_d-\q_u\right) \left(\q_e-\q_u\right)}{2} \right\}_1 \lwc{duu}{LR}[S][prst]  \,, \nn
	\delta_\mathrm{fin}^{a_\mathrm{ev}}\Bigl( \lwc{duu}{RL}[S][prst] \Bigr) &= \left\{ (1+2a_\mathrm{ev}) \frac{e^2 \left(\q_d-\q_u\right) \left(\q_e-\q_u\right)}{2} \right\}_1 \lwc{duu}{RL}[S][prst] \,, \nn
	\delta_\mathrm{fin}^{a_\mathrm{ev}}\Bigl( \lwc{udd}{LR}[S][prst] \Bigr) &= \left\{ (1+2a_\mathrm{ev}) \frac{e^2 \q_d \left(\q_u-\q_d\right)}{2} \right\}_1 \lwc{udd}{LR}[S][prst] \,, \nn
	\delta_\mathrm{fin}^{a_\mathrm{ev}}\Bigl( \lwc{dud}{RL}[S][prst] \Bigr) &= \left\{ (1+2a_\mathrm{ev}) \frac{e^2 \q_d \left(\q_u-\q_d\right)}{2} \right\}_1 \lwc{dud}{RL}[S][prst] \,.
\end{align}
Taking into account these finite shifts both in the counterterm diagrams and in the master formula~\eqref{eq:RGEMasterFormula} leads to the following $a_\mathrm{ev}$-dependent terms in the two-loop RGEs:
\begin{align}
	\label{eq:aEvSchemeChange}
	\left[ \dlwc{duu}{LR}[S][prst] \right]_2^{a_\mathrm{ev}} &= (1+2a_\mathrm{ev}) b_{0,0}^e e^4 \left(\q_d-\q_u\right) \left(\q_e-\q_u\right) \lwc{duu}{LR}[S][prst]  \,, \nn
	\left[ \dlwc{duu}{RL}[S][prst] \right]_2^{a_\mathrm{ev}} &= (1+2a_\mathrm{ev}) b_{0,0}^e e^4 \left(\q_d-\q_u\right) \left(\q_e-\q_u\right) \lwc{duu}{RL}[S][prst] \,, \nn
	\left[ \dlwc{udd}{LR}[S][prst] \right]_2^{a_\mathrm{ev}} &= (1+2a_\mathrm{ev}) b_{0,0}^e e^4 \q_d \left(\q_u-\q_d\right) \lwc{udd}{LR}[S][prst] \,, \nn
	\left[ \dlwc{dud}{RL}[S][prst] \right]_2^{a_\mathrm{ev}} &= (1+2a_\mathrm{ev}) b_{0,0}^e e^4 \q_d \left(\q_u-\q_d\right) \lwc{dud}{RL}[S][prst] \,.
\end{align}
A dependence on $e_\mathrm{ev}$ does not arise, because no structures with four Dirac matrices are generated from one-loop insertions of physical operators.

\subsection{Comparison to literature}

Our results in the chirally symmetric HV scheme are all new to the best of our knowledge. The NDR results can be compared to existing literature. For the pure QCD contribution to same-chirality operators, we find agreement with Ref.~\cite{Nihei:1994tx}. The operators with mixed chirality were written in Ref.~\cite{Nihei:1994tx} in terms of vector bilinears, whereas in our basis we use scalar operators throughout. As mentioned at the end of Sect.~\ref{sec:NDREvanescents}, the relation between these operator bases involves Fierz-evanescent terms beyond the evanescent operators used here. As we have not computed the insertions of these additional evanescent operators, the two-loop RGEs for mixed chiralities cannot be compared directly to Ref.~\cite{Nihei:1994tx}.

The RGEs in pure QCD are in fact known even at the three-loop level: in Ref.~\cite{Gracey:2012gx}, they were obtained in the Larin scheme, including finite renormalizations that restore chiral symmetry and are expected to bring the results into agreement with the NDR scheme if the same evanescent operators are used. Indeed, we find agreement of our QCD results in the NDR scheme with the two-loop contributions of Ref.~\cite{Gracey:2012gx}, both for same-chirality and mixed-chirality operators.

Very recently, the two-loop RGEs in the NDR scheme were also obtained in Ref.~\cite{Aebischer:2025hsx} as a part of the complete set of two-loop RGEs of four-fermion operators in the LEFT.
 The $n_f$-dependence is not provided there, but explicit values for different fixed numbers of flavors are given. To facilitate the comparison, we evaluate our RGEs explicitly for $n_u = 2$, $n_d = n_e = 3$, which corresponds to the LEFT between the electroweak scale and the $b$-quark threshold, see Eqs.~\eqref{eq:NDR2LoopRGEsFixedNfDeltaLplus1} and~\eqref{eq:NDR2LoopRGEsFixedNfDeltaLminus1}. After finding only partial agreement with the supplementary material of the arXiv version 1 of Ref.~\cite{Aebischer:2025hsx}, we have been in contact with the authors of Ref.~\cite{Aebischer:2025hsx}: we have been informed that after revisiting their calculation, they fully agree with our results, as will be reflected in an updated version of their work.\footnote{Comparing to arXiv version 1 of Ref.~\cite{Aebischer:2025hsx}, we found that several of the mixed QCD and QED terms of $\O(g^2e^2)$ agreed (which in our calculation are identical in the NDR and HV schemes) and also for many of the $\O(e^4)$ terms we found agreement, while the QCD contributions did not agree. Surprisingly, Ref.~\cite{Aebischer:2025hsx} did not always find the same RGEs for operators related by parity, in particular the RGEs for $\lwc{udd}{LL}[S]$ and $\lwc{udd}{RR}[S]$ differed, as did the ones for $\lwc{dud}{RL}[S]$ and $\lwc{udd}{LR}[S]$. The results for different number of flavors also indicated that the $n_f$-dependence of the QCD contribution in Ref.~\cite{Aebischer:2025hsx} could not be written in terms of $b^g_{0,0}$.}

%% file: sections/Conclusions.tex

\section{Conclusions}
\label{sec:Conclusions}

In the present work, we have calculated the two-loop RGEs for the $B$-violating sector of the LEFT at dimension six. We have obtained the results in two different schemes. The first one is our preferred scheme, the HV scheme including finite renormalizations that compensate spurious chiral-symmetry-breaking effects as well as evanescent insertions~\cite{Naterop:2023dek}. Its advantage of mathematical rigor comes at the expense of much higher computational complexity. The second scheme is the NDR scheme, which in the considered sector of the LEFT does not lead to any ill-defined Dirac traces at two loops. In NDR, we define the evanescent scheme in such a way that the canonical flavor symmetries of the operators are respected by their Wilson coefficients. A careful scheme choice preserving symmetries allows to maximally restrict the flavor structure of the RGEs at any loop order. For the differences between the two schemes, we find a simple expression~\eqref{eq:NDRvsHV}, which can be deduced from the form of the finite one-loop renormalizations in the HV scheme. While our results in the HV scheme are new, the NDR results are confirmed independently~\cite{Aebischer:2025hsx}.

Our work is one part in a series of papers establishing the complete two-loop RGEs of the LEFT up to dimension six in the HV scheme~\cite{Naterop:2023dek,Naterop:2024ydo}. The generalization of our analysis of the $B$-violating sector to the SMEFT above the electroweak scale is a natural next step and work in progress~\cite{Banik:2025inPrep}. Together with the results on the LEFT presented here, it will enable a comprehensive EFT analysis of proton-decay constraints on physics beyond the SM at NLL accuracy.

%% file: sections/RGE.tex

\section{Two-loop RGEs}
\label{sec:RGEs}

Following the conventions of Ref.~\cite{Naterop:2024ydo}, we write the RGEs in the form
\begin{equation}
	\label{eq:RGENotation}
	\dot X = \frac{d}{d\log\mu} X = \frac{1}{16\pi^2} [\dot X]_1 + \frac{1}{(16\pi^2)^2} [\dot X]_2 \, .
\end{equation}
The scheme-independent one-loop contribution to the RGEs $[\dot X]_1$ have been computed in Ref.~\cite{Jenkins:2017dyc}. Below, we provide the two-loop contribution $[\dot X]_2$ to the RGEs in both the HV and the NDR schemes. The difference between the two schemes has a simple form, given in Eq.~\eqref{eq:NDRvsHV}.

\subsection{RGEs in the HV scheme}

In our chirally symmetric HV scheme, the two-loop contribution to the RGEs in the $\Delta B = \Delta L = 1$ sector read
\begin{align}
	\left[ \dlwc{udd}{LL}[S][prst] \right]_2 &= \lwc{udd}{LL}[S][prst] \biggl(\left(\frac{1}{6} b_{0,0}^e \left(20 \q_d^2+9 \q_u^2 - 2 \q_d \q_u\right)-8 \q_d \q_u^3-32 \q_d^2 \q_u^2-8 \q_d^3 \q_u-27 \q_d^4-\frac{3 \q_u^4}{2}\right) e^4 \nn*
		&\qquad\qquad + \frac{4}{3} \left(16 \q_d \q_u+10 \q_d^2+\q_u^2\right) g^2 e^2 + 6 \left(b_{0,0}^g-2\right) g^4 \biggr) \nn*
		&\quad + \lwc{udd}{LL}[S][psrt] \left(\q_u-\q_d\right) \left(\q_d \left(\frac{2}{3} b_{0,0}^e - 8 \left(2 \q_d \q_u+3 \q_d^2+\q_u^2\right)\right)e^4 + \frac{16}{3} \left(\q_u-\q_d\right) g^2 e^2 \right) \,, \\
	\left[ \dlwc{duu}{LL}[S][prst] \right]_2 &= \lwc{duu}{LL}[S][prst] \biggl( \biggl(\frac{1}{6} b_{0,0}^e \left( 2 \q_d \q_e+9 \q_d^2+9 \q_e^2+20 \q_u^2 - 2 \q_u \left(\q_d+\q_e\right) \right) \nn*
		&\qquad\qquad\quad - 8 \q_u^3 \left(\q_d+\q_e\right) - 32 \q_u^2 \left(\q_d^2+\q_e^2\right) - 8 \q_u \left(\q_d^3+\q_e^3\right) \nn*
		&\qquad\qquad\quad - \frac{1}{2} \left(8 \q_d^3 \q_e + 32 \q_d^2 \q_e^2 + 8 \q_d \q_e^3 + 3 \q_d^4 + 3 \q_e^4\right)-27 \q_u^4 \biggr) e^4 \nn*
		&\qquad\qquad\quad + \frac{4}{3} \left(-4 \q_d \left(\q_e-4 \q_u\right)+\q_d^2+2 \q_u \left(5 \q_u-4 \q_e\right)\right) g^2 e^2  + 6 \left(b_{0,0}^g-2\right) g^4 \biggr) \nn*
	   	&\quad + \lwc{duu}{LL}[S][psrt] \left(\q_d-\q_u\right) \biggl( \left(\q_u-\q_e\right) \left(\frac{2}{3} b_{0,0}^e - 8 \left(2 \q_u \left(\q_d+\q_e\right)+\q_d \q_e+\q_d^2+\q_e^2+3 \q_u^2\right)\right) e^4 \nn*
		&\qquad\qquad\qquad\qquad + \frac{16}{3} \left(\q_d+2 \q_e-\q_u\right) g^2 e^2 \biggr) \, ,  \\
	\left[ \dlwc{uud}{LR}[S][prst] \right]_2 &= \lwc{uud}{LR}[S][prst] \biggl( \frac{3}{2} \Bigl( \left(\q_d^2+\q_e^2+2 \q_u^2\right) b_{0,0}^e \nn*
		&\qquad\qquad\quad - \left(\q_d^4+8 \q_e \q_d^3+16 \q_e^2 \q_d^2+8 \q_e \left(\q_e^2+\q_u^2\right) \q_d+\q_e^4+34 \q_u^4\right)\Bigr) e^4 \nn*
		&\qquad\qquad - 4 \left(\q_d^2+4 \q_e \q_d-2 \q_u \left(\q_e+\q_u\right)\right) g^2 e^2 + \left(6 b_{0,0}^g-\frac{76}{3}\right) g^4 \biggr) \,, \\
	\left[ \dlwc{duu}{LR}[S][prst] \right]_2 &= \lwc{duu}{LR}[S][prst] \biggl(\frac{3}{2} \Bigl(\left(\q_d^2+\q_e^2+2 \q_u^2\right) b_{0,0}^e-12 \left(\q_d+\q_e\right) \q_u^3-\left(\q_d^2-\q_e^2\right)^2-14 \left(\q_d^2+\q_e^2\right) \q_u^2 \nn*
		&\qquad\qquad\quad - 4 \left(\q_d+\q_e\right) \left(2 \q_d^2-\q_e \q_d+2 \q_e^2\right) \q_u \Bigr) e^4 \nn*
		&\qquad\qquad + 4 \left(\q_d^2+\q_e \q_d-3 \q_e \q_u\right) g^2 e^2 + \left(6 b_{0,0}^g-\frac{76}{3}\right) g^4 \biggr) \,, \\
	\left[ \dlwc{uud}{RL}[S][prst] \right]_2 &= \lwc{uud}{RL}[S][prst] \biggl( \frac{3}{2} \Bigl( \left(\q_d^2+\q_e^2+2 \q_u^2\right) b_{0,0}^e \nn*
		&\qquad\qquad\quad - \left(\q_d^4+8 \q_e \q_d^3+16 \q_e^2 \q_d^2+8 \q_e \left(\q_e^2+\q_u^2\right) \q_d+\q_e^4+34 \q_u^4\right) \Bigr) e^4 \nn*
		&\qquad\qquad - 4 \left(\q_d^2+4 \q_e \q_d-2 \q_u \left(\q_e+\q_u\right)\right) g^2 e^2 + \left(6 b_{0,0}^g-\frac{76}{3}\right) g^4 \biggr) \,, \\
	\left[ \dlwc{duu}{RL}[S][prst] \right]_2 &= \lwc{duu}{RL}[S][prst] \biggl( \frac{3}{2} \Bigl( \left(\q_d^2+\q_e^2+2 \q_u^2\right) b_{0,0}^e-12 \left(\q_d+\q_e\right) \q_u^3-\left(\q_d^2-\q_e^2\right)^2-14 \left(\q_d^2+\q_e^2\right) \q_u^2 \nn*
		&\qquad\qquad\quad - 4 \left(\q_d+\q_e\right) \left(2 \q_d^2-\q_e \q_d+2 \q_e^2\right) \q_u \Bigr) e^4 \nn*
		&\qquad\qquad + 4 \left(\q_d^2+\q_e \q_d-3 \q_e \q_u\right) g^2 e^2 + \left(6 b_{0,0}^g-\frac{76}{3}\right) g^4 \biggr) \,, \\
	\left[ \dlwc{dud}{RL}[S][prst] \right]_2 &= \lwc{dud}{RL}[S][prst] \biggl(\frac{3}{2} \left(\left(2 \q_d^2+\q_u^2\right) b_{0,0}^e-\q_u \left(6 \q_d+\q_u\right) \left(2 \q_d^2+2 \q_u \q_d+\q_u^2\right)\right) e^4 \nn*
		&\qquad\qquad + 4 \q_u^2 g^2 e^2 + \left(6 b_{0,0}^g-\frac{76}{3}\right) g^4 \biggr) \,, \\
	\left[ \dlwc{ddu}{RL}[S][prst] \right]_2 &= \lwc{ddu}{RL}[S][prst] \biggl(\frac{3}{2} \left(\left(2 \q_d^2+\q_u^2\right) b_{0,0}^e - 34 \q_d^4-\q_u^4\right) e^4 \nn*
		&\qquad\qquad + \left(8 \q_d^2-4 \q_u^2\right) g^2 e^2 + \left(6 b_{0,0}^g-\frac{76}{3}\right) g^4 \biggr) \,, \\
	\left[ \dlwc{duu}{RR}[S][prst] \right]_2 &= \lwc{duu}{RR}[S][prst] \biggl( \biggl(\frac{1}{6} b_{0,0}^e \left( 2 \q_d \q_e+9 \q_d^2+9 \q_e^2+20 \q_u^2 - 2 \q_u \left(\q_d+\q_e\right) \right) \nn*
		&\qquad\qquad\quad - 8 \q_u^3 \left(\q_d+\q_e\right) - 32 \q_u^2 \left(\q_d^2+\q_e^2\right) - 8 \q_u \left(\q_d^3+\q_e^3\right) \nn*
		&\qquad\qquad\quad - \frac{1}{2} \left(8 \q_d^3 \q_e + 32 \q_d^2 \q_e^2 + 8 \q_d \q_e^3 + 3 \q_d^4 + 3 \q_e^4\right)-27 \q_u^4 \biggr) e^4 \nn*
		&\qquad\qquad\quad + \frac{4}{3} \left(-4 \q_d \left(\q_e-4 \q_u\right)+\q_d^2+2 \q_u \left(5 \q_u-4 \q_e\right)\right) g^2 e^2  + 6 \left(b_{0,0}^g-2\right) g^4 \biggr) \nn*
	   	&\quad + \lwc{duu}{RR}[S][psrt] \left(\q_d-\q_u\right) \biggl( \left(\q_u-\q_e\right) \left(\frac{2}{3} b_{0,0}^e - 8 \left(2 \q_u \left(\q_d+\q_e\right)+\q_d \q_e+\q_d^2+\q_e^2+3 \q_u^2\right)\right) e^4 \nn*
		&\qquad\qquad\qquad\qquad + \frac{16}{3} \left(\q_d+2 \q_e-\q_u\right) g^2 e^2 \biggr) \, .
\end{align}
For the $\Delta B = -\Delta L = 1$ sector, we obtain
\begin{align}
	\left[ \dlwc{ddd}{LL}[S][prst] \right]_2 &= \lwc{ddd}{LL}[S][prst] \biggl(\frac{3}{2} \Bigl(\left(3 \q_d^2+\q_e^2\right) b_{0,0}^e-51 \q_d^4-\q_e^4+8 \q_d \q_e^3-32 \q_d^2 \q_e^2+8 \q_d^3 \q_e\Bigr) e^4 \nn*
		&\qquad\qquad + 4 \q_d \left(9 \q_d+4 \q_e\right) g^2 e^2 + 6 \left(b_{0,0}^g-2\right) g^4 \biggr) \,, \\
	\left[ \dlwc{udd}{LR}[S][prst] \right]_2 &= \lwc{udd}{LR}[S][prst] \biggl(\frac{3}{2} \Bigl(\left(2 \q_d^2+\q_u^2\right) b_{0,0}^e-\q_u \left(6 \q_d+\q_u\right) \left(2 \q_d^2+2 \q_u \q_d+\q_u^2\right)\Bigr) e^4 \nn*
		&\qquad\qquad + 4 \q_u^2 g^2 e^2 + \left(6 b_{0,0}^g-\frac{76}{3}\right) g^4 \biggr) \,, \\
	\left[ \dlwc{ddu}{LR}[S][prst] \right]_2 &= \lwc{ddu}{LR}[S][prst] \biggl(\frac{3}{2} \Bigl(\left(2 \q_d^2+\q_u^2\right) b_{0,0}^e-34 \q_d^4-\q_u^4\Bigr) e^4 \nn*
		&\qquad\qquad + \left(8 \q_d^2-4 \q_u^2\right) g^2 e^2 + \left(6 b_{0,0}^g-\frac{76}{3}\right) g^4 \biggr) \,, \\
	\left[ \dlwc{ddd}{LR}[S][prst] \right]_2 &= \lwc{ddd}{LR}[S][prst] \biggl(\frac{3}{2} \Bigl(\left(3 \q_d^2+\q_e^2\right) b_{0,0}^e-35 \q_d^4-\q_e^4+8 \q_d \q_e^3-16 \q_d^2 \q_e^2+16 \q_d^3 \q_e\Bigr) e^4 \nn*
		&\qquad\qquad + 4 \q_d \left(\q_d+2 \q_e\right) g^2 e^2 + \left(6 b_{0,0}^g-\frac{76}{3}\right) g^4 \biggr) \,, \\
	\left[ \dlwc{ddd}{RL}[S][prst] \right]_2 &= \lwc{ddd}{RL}[S][prst] \biggl(\frac{3}{2} \Bigl(\left(3 \q_d^2+\q_e^2\right) b_{0,0}^e-35 \q_d^4-\q_e^4+8 \q_d \q_e^3-16 \q_d^2 \q_e^2+16 \q_d^3 \q_e\Bigr) e^4 \nn*
		&\qquad\qquad + 4 \q_d \left(\q_d+2 \q_e\right) g^2 e^2 + \left(6 b_{0,0}^g-\frac{76}{3}\right) g^4 \biggr) \,, \\
	\left[ \dlwc{udd}{RR}[S][prst] \right]_2 &=  \lwc{udd}{RR}[S][prst] \biggl(\Bigl(\frac{1}{6} \left(20 \q_d^2-2 \q_u \q_d+9 \q_u^2\right) b_{0,0}^e-27 \q_d^4-\frac{3 \q_u^4}{2}-8 \q_d \q_u^3-32 \q_d^2 \q_u^2-8 \q_d^3 \q_u\Bigr) e^4 \nn*
		&\qquad\qquad +\frac{4}{3} \left(10 \q_d^2+16 \q_u \q_d+\q_u^2\right) g^2 e^2 + 6 \left(b_{0,0}^g-2\right) g^4 \biggr) \nn*
		&\quad + \lwc{udd}{RR}[S][ptsr] (\q_d-\q_u) \biggl(\Bigl(8 \q_d \left(3 \q_d^2+2 \q_u \q_d+\q_u^2\right)-\frac{2}{3} \q_d b_{0,0}^e\Bigr) e^4+\frac{16}{3} \left(\q_d-\q_u\right) g^2 e^2 \biggr) \,, \\
	\left[ \dlwc{ddd}{RR}[S][prst] \right]_2 &= \lwc{ddd}{RR}[S][prst] \biggl(\frac{3}{2} \Bigl(\left(3 \q_d^2+\q_e^2\right) b_{0,0}^e-51 \q_d^4-\q_e^4+8 \q_d \q_e^3-32 \q_d^2 \q_e^2+8 \q_d^3 \q_e\Bigr) e^4 \nn*
		&\qquad\qquad + 4 \q_d \left(9 \q_d+4 \q_e\right) g^2 e^2 + 6 \left(b_{0,0}^g-2\right) g^4 \biggr) \,.
\end{align}

\subsection{RGEs in the NDR scheme}

In the NDR scheme, the two-loop contribution to the RGEs in the $\Delta B = \Delta L = 1$ sector read for $a_\mathrm{ev} = -1/2$
\begin{align}
	\left[ \dlwc{udd}{LL}[S][prst] \right]_2 &= \lwc{udd}{LL}[S][prst] \biggl( \Bigl(\frac{1}{6}b_{0,0}^e \left(20 \q_d^2+9 \q_u^2 + 46 \q_d \q_u\right) - 8 \q_d^3 \q_u - 32 \q_d^2 \q_u^2 - 8 \q_d \q_u^3 - 27 \q_d^4 - \frac{3 \q_u^4}{2} \Bigr) e^4 \nn*
		&\qquad\qquad + \frac{4}{3} \left(16 \q_d \q_u+10 \q_d^2+\q_u^2\right) g^2 e^2 + \frac{2}{3} \left(b_{0,0}^g-18\right) g^4 \biggr) \nn*
   		&\quad+ \lwc{udd}{LL}[S][psrt] \left(\q_u-\q_d\right) \left(  \q_d \Bigl( \frac{2}{3} b_{0,0}^e - 8 \left(2 \q_d \q_u+3 \q_d^2+\q_u^2\right)\Bigr) e^4 + \frac{16}{3} \left(\q_u-\q_d\right) g^2 e^2 \right) \, , \\
	\left[ \dlwc{duu}{LL}[S][prst] \right]_2 &= \lwc{duu}{LL}[S][prst] \biggl( \biggl(\frac{1}{6} b_{0,0}^e \left( 2 \q_d \q_e+9 \q_d^2+9 \q_e^2+20 \q_u^2 + 46 \q_u \left(\q_d+\q_e\right) \right) \nn*
		&\qquad\qquad\quad - 8 \q_u^3 \left(\q_d+\q_e\right) - 32 \q_u^2 \left(\q_d^2+\q_e^2\right) - 8 \q_u \left(\q_d^3+\q_e^3\right) \nn*
		&\qquad\qquad\quad - \frac{1}{2} \left(8 \q_d^3 \q_e + 32 \q_d^2 \q_e^2 + 8 \q_d \q_e^3 + 3 \q_d^4 + 3 \q_e^4\right)-27 \q_u^4 \biggr) e^4 \nn*
		&\qquad\qquad\quad + \frac{4}{3} \left(-4 \q_d \left(\q_e-4 \q_u\right)+\q_d^2+2 \q_u \left(5 \q_u-4 \q_e\right)\right) g^2 e^2  + \frac{2}{3} \left(b_{0,0}^g-18\right) g^4 \biggr) \nn*
	   	&\quad + \lwc{duu}{LL}[S][psrt] \left(\q_d-\q_u\right) \biggl( \left(\q_u-\q_e\right) \left(\frac{2}{3} b_{0,0}^e - 8 \left(2 \q_u \left(\q_d+\q_e\right)+\q_d \q_e+\q_d^2+\q_e^2+3 \q_u^2\right)\right) e^4 \nn*
		&\qquad\qquad\qquad\qquad + \frac{16}{3} \left(\q_d+2 \q_e-\q_u\right) g^2 e^2 \biggr) \, ,  \\
	\left[ \dlwc{uud}{LR}[S][prst] \right]_2 &= \lwc{uud}{LR}[S][prst] \biggl( \frac{1}{2} \Bigl(\left(3 \q_d^2+16 \q_e \q_d+3 \q_e^2+22 \q_u^2\right) b_{0,0}^e \nn*
		&\qquad\qquad\quad - 3 \left(\q_d^4+8 \q_e \q_d^3+16 \q_e^2 \q_d^2+8 \q_e \left(\q_e^2+\q_u^2\right) \q_d+\q_e^4+34 \q_u^4\right)\Bigr) e^4 \nn*
		&\qquad\qquad - 4 \left(\q_d^2+4 \q_e \q_d-2 \q_u \left(\q_e+\q_u\right)\right) g^2 e^2 + \frac{2}{3} \left(b_{0,0}^g-38\right) g^4 \biggr) \,, \\
	\left[ \dlwc{duu}{LR}[S][prst] \right]_2 &= \lwc{duu}{LR}[S][prst] \biggl(\Bigl(\Bigl( 8\q_u(\q_d+\q_e) + \frac{3}{2} \left(\q_d^2+\q_e^2\right) + 3 \q_u^2\Bigr) b_{0,0}^e-18 \left(\q_d+\q_e\right) \q_u^3- \frac{3}{2} \left(\q_d^2-\q_e^2\right)^2 \nn*
		&\qquad\qquad\quad - 21 \left(\q_d^2+\q_e^2\right) \q_u^2 - 6 \left(\q_d+\q_e\right) \left(2 \q_d^2-\q_e \q_d+2 \q_e^2\right) \q_u \Bigr) e^4 \nn*
		&\qquad\qquad + 4 \left(\q_d^2+\q_e \q_d-3 \q_e \q_u\right) g^2 e^2 + \frac{2}{3} \left(b_{0,0}^g-38\right) g^4 \biggr) \,, \\
	\left[ \dlwc{uud}{RL}[S][prst] \right]_2 &= \lwc{uud}{RL}[S][prst] \biggl( \frac{1}{2} \Bigl( \left(3 \q_d^2 + 16\q_e\q_d + 3 \q_e^2 + 22 \q_u^2\right) b_{0,0}^e \nn*
		&\qquad\qquad\quad - 3 \left(\q_d^4+8 \q_e \q_d^3+16 \q_e^2 \q_d^2+8 \q_e \left(\q_e^2+\q_u^2\right) \q_d+\q_e^4+34 \q_u^4\right) \Bigr) e^4 \nn*
		&\qquad\qquad - 4 \left(\q_d^2+4 \q_e \q_d-2 \q_u \left(\q_e+\q_u\right)\right) g^2 e^2 + \frac{2}{3} \left(b_{0,0}^g-38\right) g^4 \biggr) \,, \\
	\left[ \dlwc{duu}{RL}[S][prst] \right]_2 &= \lwc{duu}{RL}[S][prst] \biggl(\Bigl(\Bigl( 8\q_u(\q_d+\q_e) + \frac{3}{2} \left(\q_d^2+\q_e^2\right) + 3 \q_u^2\Bigr) b_{0,0}^e-18 \left(\q_d+\q_e\right) \q_u^3- \frac{3}{2} \left(\q_d^2-\q_e^2\right)^2 \nn*
		&\qquad\qquad\quad - 21 \left(\q_d^2+\q_e^2\right) \q_u^2 - 6 \left(\q_d+\q_e\right) \left(2 \q_d^2-\q_e \q_d+2 \q_e^2\right) \q_u \Bigr) e^4 \nn*
		&\qquad\qquad + 4 \left(\q_d^2+\q_e \q_d-3 \q_e \q_u\right) g^2 e^2 + \frac{2}{3} \left(b_{0,0}^g-38\right) g^4 \biggr) \,, \\
	\left[ \dlwc{dud}{RL}[S][prst] \right]_2 &= \lwc{dud}{RL}[S][prst] \biggl( \left(\Bigl(8 \q_d \q_u+3 \q_d^2+\frac{3 \q_u^2}{2}\Bigr)b_{0,0}^e - \frac{3}{2}\q_u \left(6 \q_d+\q_u\right) \left(2 \q_d^2+2 \q_u \q_d+\q_u^2\right)\right) e^4 \nn*
		&\qquad\qquad + 4 \q_u^2 g^2 e^2 + \frac{2}{3} \left(b_{0,0}^g-38\right) g^4 \biggr) \,, \\
	\left[ \dlwc{ddu}{RL}[S][prst] \right]_2 &= \lwc{ddu}{RL}[S][prst] \biggl(\Bigl(\Bigl(11 \q_d^2+\frac{3 \q_u^2}{2}\Bigr) b_{0,0}^e-\frac{3}{2} \left(34 \q_d^4+\q_u^4\right)\Bigr) e^4 \nn*
		&\qquad\qquad + 4 \left(2 \q_d^2- \q_u^2\right) g^2 e^2 + \frac{2}{3} \left(b_{0,0}^g-38\right) g^4 \biggr) \,, \\
	\left[ \dlwc{duu}{RR}[S][prst] \right]_2 &= \lwc{duu}{RR}[S][prst] \biggl( \biggl(\frac{1}{6} b_{0,0}^e \left( 2 \q_d \q_e+9 \q_d^2+9 \q_e^2+20 \q_u^2 + 46 \q_u \left(\q_d+\q_e\right) \right) \nn*
		&\qquad\qquad\quad - 8 \q_u^3 \left(\q_d+\q_e\right) - 32 \q_u^2 \left(\q_d^2+\q_e^2\right) - 8 \q_u \left(\q_d^3+\q_e^3\right) \nn*
		&\qquad\qquad\quad - \frac{1}{2} \left(8 \q_d^3 \q_e + 32 \q_d^2 \q_e^2 + 8 \q_d \q_e^3 + 3 \q_d^4 + 3 \q_e^4\right)-27 \q_u^4 \biggr) e^4 \nn*
		&\qquad\qquad\quad + \frac{4}{3} \left(-4 \q_d \left(\q_e-4 \q_u\right)+\q_d^2+2 \q_u \left(5 \q_u-4 \q_e\right)\right) g^2 e^2  + \frac{2}{3} \left(b_{0,0}^g-18\right) g^4 \biggr) \nn*
	   	&\quad + \lwc{duu}{RR}[S][psrt] \left(\q_d-\q_u\right) \biggl( \left(\q_u-\q_e\right) \left(\frac{2}{3} b_{0,0}^e - 8 \left(2 \q_u \left(\q_d+\q_e\right)+\q_d \q_e+\q_d^2+\q_e^2+3 \q_u^2\right)\right) e^4 \nn*
		&\qquad\qquad\qquad\qquad + \frac{16}{3} \left(\q_d+2 \q_e-\q_u\right) g^2 e^2 \biggr) \, ,
\end{align}
whereas for the $\Delta B = -\Delta L = 1$ sector, we obtain
\begin{align}
	\left[ \dlwc{ddd}{LL}[S][prst] \right]_2 &= \lwc{ddd}{LL}[S][prst] \biggl(\frac{1}{2} \left(\left(25 \q_d^2-16 \q_e \q_d+3 \q_e^2\right) b_{0,0}^e-3 \left(51 \q_d^4-8 \q_e \q_d^3+32 \q_e^2 \q_d^2-8 \q_e^3 \q_d+\q_e^4\right)\right) e^4 \nn*
		&\qquad\qquad + 4 \q_d \left(9 \q_d+4 \q_e\right) g^2 e^2+\frac{2}{3} \left(b_{0,0}^g-18\right) g^4 \biggr) \,, \\
	\left[ \dlwc{udd}{LR}[S][prst] \right]_2 &= \lwc{udd}{LR}[S][prst] \biggl( \left(\Bigl(8 \q_d \q_u+3 \q_d^2+\frac{3 \q_u^2}{2}\Bigr)b_{0,0}^e - \frac{3}{2}\q_u \left(6 \q_d+\q_u\right) \left(2 \q_d^2+2 \q_u \q_d+\q_u^2\right)\right) e^4 \nn*
		&\qquad\qquad + 4 \q_u^2 g^2 e^2 + \frac{2}{3} \left(b_{0,0}^g-38\right) g^4 \biggr) \,, \\
	\left[ \dlwc{ddu}{LR}[S][prst] \right]_2 &= \lwc{ddu}{LR}[S][prst] \biggl(\Bigl(\Bigl(11 \q_d^2+\frac{3 \q_u^2}{2}\Bigr) b_{0,0}^e-\frac{3}{2} \left(34 \q_d^4+\q_u^4\right)\Bigr) e^4 \nn*
		&\qquad\qquad + 4 \left(2 \q_d^2- \q_u^2\right) g^2 e^2 + \frac{2}{3} \left(b_{0,0}^g-38\right) g^4 \biggr) \,, \\
	\left[ \dlwc{ddd}{LR}[S][prst] \right]_2 &= \lwc{ddd}{LR}[S][prst] \biggl(\frac{1}{2} \left(\left(25 \q_d^2-16 \q_e \q_d+3 \q_e^2\right) b_{0,0}^e-3 \left(35 \q_d^4-16 \q_e \q_d^3+16 \q_e^2 \q_d^2-8 \q_e^3 \q_d+\q_e^4\right)\right) e^4 \nn*
		&\qquad\qquad + 4 \q_d \left(\q_d+2 \q_e\right) g^2 e^2+\frac{2}{3} \left(b_{0,0}^g-38\right) g^4 \biggr) \,, \\
	\left[ \dlwc{ddd}{RL}[S][prst] \right]_2 &= \lwc{ddd}{RL}[S][prst] \biggl(\frac{1}{2} \left(\left(25 \q_d^2-16 \q_e \q_d+3 \q_e^2\right) b_{0,0}^e-3 \left(35 \q_d^4-16 \q_e \q_d^3+16 \q_e^2 \q_d^2-8 \q_e^3 \q_d+\q_e^4\right)\right) e^4 \nn*
		&\qquad\qquad + 4 \q_d \left(\q_d+2 \q_e\right) g^2 e^2 + \frac{2}{3} \left(b_{0,0}^g-38\right) g^4 \biggr) \,, \\
	\left[ \dlwc{udd}{RR}[S][prst] \right]_2 &= \lwc{udd}{RR}[S][prst] \biggl( \Bigl(\frac{1}{6}b_{0,0}^e \left(20 \q_d^2+9 \q_u^2 + 46 \q_d \q_u\right) - 8 \q_d^3 \q_u - 32 \q_d^2 \q_u^2 - 8 \q_d \q_u^3 - 27 \q_d^4 - \frac{3 \q_u^4}{2} \Bigr) e^4 \nn*
		&\qquad\qquad + \frac{4}{3} \left(16 \q_d \q_u+10 \q_d^2+\q_u^2\right) g^2 e^2 + \frac{2}{3} \left(b_{0,0}^g-18\right) g^4 \biggr) \nn*
   		&\quad+ \lwc{udd}{RR}[S][ptsr] \left(\q_u-\q_d\right) \left(  \q_d \Bigl( \frac{2}{3} b_{0,0}^e - 8 \left(2 \q_d \q_u+3 \q_d^2+\q_u^2\right)\Bigr) e^4 + \frac{16}{3} \left(\q_u-\q_d\right) g^2 e^2 \right) \, , \\
	\left[ \dlwc{ddd}{RR}[S][prst] \right]_2 &= \lwc{ddd}{RR}[S][prst] \biggl(\frac{1}{2} \left(\left(25 \q_d^2-16 \q_e \q_d+3 \q_e^2\right) b_{0,0}^e-3 \left(51 \q_d^4-8 \q_e \q_d^3+32 \q_e^2 \q_d^2-8 \q_e^3 \q_d+\q_e^4\right)\right) e^4 \nn*
		&\qquad\qquad + 4 \q_d \left(9 \q_d+4 \q_e\right) g^2 e^2 + \frac{2}{3} \left(b_{0,0}^g-18\right) g^4 \biggr) \,.
\end{align}
The scheme change to different values of $a_\mathrm{ev}$ is given in Eq.~\eqref{eq:aEvSchemeChange}.
For convenience and to facilitate comparison to Ref.~\cite{Aebischer:2025hsx}, we provide the two-loop coefficients of the RGEs in the NDR scheme for the LEFT between the $b$-quark threshold and the electroweak scale, i.e., for $n_u = 2$, $n_d = n_e = 3$, including the dependence on the scheme parameter $a_\mathrm{ev}$. For $\Delta B = \Delta L = 1$, they are given by
\begin{align}
	\label{eq:NDR2LoopRGEsFixedNfDeltaLplus1}
	\left[ \dlwc{udd}{LL}[S][prst] \right]_2 &= \left({\frac{127}{27} e^4 - \frac{8 g^2 e^2}{3}} -\frac{62 g^4}{9}\right) \lwc{udd}{LL}[S][prst] + \left(\frac{232 e^4}{81} + {\frac{16 g^2 e^2}{3}}\right) \lwc{udd}{LL}[S][psrt] \,, \nn
	\left[ \dlwc{duu}{LL}[S][prst] \right]_2 &= \left({\frac{1165}{81} e^4+ \frac{20 g^2 e^2}{3}}-\frac{62 g^4}{9}\right) \lwc{duu}{LL}[S][prst] + \left(\frac{1880 e^4}{81}+{16 g^2 e^2}\right) \lwc{duu}{LL}[S][psrt] \,, \nn
	\left[ \dlwc{uud}{LR}[S][prst] \right]_2 &= \left({-\frac{8299 e^4}{81}-\frac{68 g^2 e^2}{9}}-\frac{182 g^4}{9}\right) \lwc{uud}{LR}[S][prst] \,, \nn
	\left[ \dlwc{duu}{LR}[S][prst] \right]_2 &= \left({-\frac{8(300a_\mathrm{ev}-277)}{81} e^4+\frac{88 g^2 e^2}{9}}-\frac{182 g^4}{9}\right) \lwc{duu}{LR}[S][prst] \,, \nn
	\left[ \dlwc{uud}{RL}[S][prst] \right]_2 &= \left({-\frac{8299 e^4}{81}-\frac{68 g^2 e^2}{9}}-\frac{182 g^4}{9}\right) \lwc{uud}{RL}[S][prst] \,, \nn
	\left[ \dlwc{duu}{RL}[S][prst] \right]_2 &= \left({-\frac{8(300a_\mathrm{ev}-277)}{81} e^4+\frac{88 g^2 e^2}{9}}-\frac{182 g^4}{9}\right) \lwc{duu}{RL}[S][prst] \,, \nn
	\left[ \dlwc{dud}{RL}[S][prst] \right]_2 &= \left({\frac{8(60a_\mathrm{ev}+103)}{81} e^4+\frac{16 g^2 e^2}{9}}-\frac{182 g^4}{9}\right) \lwc{dud}{RL}[S][prst] \,, \nn
	\left[ \dlwc{ddu}{RL}[S][prst] \right]_2 &= \left({-\frac{1435 e^4}{81}-\frac{8 g^2 e^2}{9}}-\frac{182 g^4}{9}\right) \lwc{ddu}{RL}[S][prst] \,, \nn
	\left[ \dlwc{duu}{RR}[S][prst] \right]_2 &= \left({\frac{1165}{81} e^4+\frac{20 g^2 e^2}{3}}-\frac{62 g^4}{9}\right) \lwc{duu}{RR}[S][prst] + \left(\frac{1880 e^4}{81}+{16 g^2 e^2}\right) \lwc{duu}{RR}[S][psrt] \,.
\end{align}
For $\Delta B = - \Delta L = 1$, we find
\begin{align}
	\label{eq:NDR2LoopRGEsFixedNfDeltaLminus1}
	\left[ \dlwc{ddd}{LL}[S][prst] \right]_2 &= \left(-\frac{430 e^4}{81}+\frac{28 g^2 e^2}{3}-\frac{62 g^4}{9}\right) \lwc{ddd}{LL}[S][prst] \,, \nn
	\left[ \dlwc{udd}{LR}[S][prst] \right]_2 &= \left(\frac{8(60a_\mathrm{ev}+103)}{81} e^4+\frac{16 g^2 e^2}{9}-\frac{182 g^4}{9}\right) \lwc{udd}{LR}[S][prst] \,, \nn
	\left[ \dlwc{ddu}{LR}[S][prst] \right]_2 &= \left({-\frac{1435 e^4}{81}-\frac{8 g^2 e^2}{9}}-\frac{182 g^4}{9}\right) \lwc{ddu}{LR}[S][prst] \,, \nn
	\left[ \dlwc{ddd}{LR}[S][prst] \right]_2 &= \left(-\frac{154 e^4}{81}+\frac{28 g^2 e^2}{9}-\frac{182 g^4}{9}\right) \lwc{ddd}{LR}[S][prst] \,, \nn
	\left[ \dlwc{ddd}{RL}[S][prst] \right]_2 &= \left(-\frac{154 e^4}{81}+\frac{28 g^2 e^2}{9}-\frac{182 g^4}{9}\right) \lwc{ddd}{RL}[S][prst] \,, \nn
	\left[ \dlwc{udd}{RR}[S][prst] \right]_2 &= \left(\frac{127}{27} e^4-\frac{8 g^2 e^2}{3}-\frac{62 g^4}{9}\right) \lwc{udd}{RR}[S][prst] + \left(\frac{232 e^4}{81}+\frac{16 g^2 e^2}{3}\right) \lwc{udd}{RR}[S][ptsr] \, \nn
	\left[ \dlwc{ddd}{RR}[S][prst] \right]_2 &= \left(-\frac{430 e^4}{81}+\frac{28 g^2 e^2}{3}-\frac{62 g^4}{9}\right) \lwc{ddd}{RR}[S][prst] \,.
\end{align}